\documentclass[12pt]{article}
\usepackage[reqno]{amsmath}
\usepackage{epsfig}
\usepackage{array}
\usepackage{float}
\usepackage{bbm}
\usepackage{cite}
\usepackage{a4}
\usepackage{a4wide}
\usepackage{wasysym}
\def\ra{\rightarrow}
\def\be{\begin{equation}}
\def\ee{\end{equation}}
\newcommand{\beq}{\begin{equation}}
\newcommand{\eeq}{\end{equation}}
\def\gs{\mathrel{
   \rlap{\raise 0.511ex \hbox{$>$}}{\lower 0.511ex \hbox{$\sim$}}}}
\def\ls{\mathrel{
   \rlap{\raise 0.511ex \hbox{$<$}}{\lower 0.511ex \hbox{$\sim$}}}}

\newcommand{\ba}{\begin{array}{c}}
\newcommand{\baz}{\begin{array}{cc}}
\newcommand{\bad}{\begin{array}{ccc}}
\newcommand{\bav}{\begin{array}{cccc}}
\newcommand{\bea}{\begin{equation} \begin{array}{c}}
\newcommand{\eea}{ \end{array} \end{equation}}
\newcommand{\ea}{\end{array}}

\newcommand{\dms}{\mbox{$\Delta m^2_{\odot}$}}
\newcommand{\dma}{\mbox{$\Delta m^2_{\rm A}$}}
\newcommand{\dmsol}{\mbox{$\Delta m^2_{\odot}$}}
\def\ra{\rightarrow} 
\def\gtap{\mathrel{ \rlap{\raise 0.511ex \hbox{$>$}}{\lower 0.511ex
   \hbox{$\sim$}}}} 
\def\ltap{\mathrel{ \rlap{\raise 0.511ex
   \hbox{$<$}}{\lower 0.511ex \hbox{$\sim$}}}}
   \newcommand{\deltaatm}{\mbox{$\Delta m^2_{31}$}}
   \newcommand{\deltasol}{\mbox{$ \Delta m^2_{21}$}}
   
   \newcommand{\betabeta}{\mbox{$(\beta \beta)_{0 \nu}$}}

  \newcommand{\meff}{\mbox{$ \left|< \! m \! >
            \right|$}}
  \newcommand{\mefff}{\mbox{$ < \! m \! > $}}

   \newcommand{\hbeta}{$\mbox{}^3 {\rm H}$ $\beta$-decay }

\newcommand{\pmns}{\mbox{$ U_{\rm PMNS}$}}

\def\eg{\hbox{\it e.g.}{}}
\def\etc{\hbox{\it etc}{}}

\begin{document}

\hfill{Ref.\ SISSA 26/2006/EP} 
\rightline{May 2006}
\rightline{hep-ph/0605151}

\vspace{0.5cm}

\begin{center}

{\bf \large{Charged Lepton Decays $l_i \rightarrow l_j + \gamma$,
Leptogenesis CP-Violating Parameters and Majorana Phases 
}}

\vspace{0.5cm}

S.~T.~Petcov %$^a$
\footnote{Also at: Institute of 
Nuclear Research and Nuclear Energy,
Bulgarian Academy of Sciences, 1784 Sofia, Bulgaria.}
and  T.~Shindou 

\vspace{0.2cm}

{\em Scuola Internazionale Superiore di Studi Avanzati, and \\}
% I-34014 Trieste, Italy\\ }
{\em % $^{c)}$
Istituto Nazionale di Fisica Nucleare, I-34014 Trieste, Italy\\ }

\end{center}

\vspace{0.5cm}

\renewcommand{\thefootnote}{\arabic{footnote}}
\setcounter{footnote}{0}
%
% \begin{abstract}
%
% \date{}
% \maketitle
% \thispagestyle{empty}
\begin{abstract}
\noindent 
We analyse the dependence of  the 
rates of the LFV charged lepton decays  
$\mu \to e + \gamma$, $\tau \to e + \gamma$,
$\tau \to \mu + \gamma$ ($l_i \to l_j + \gamma$) 
and their ratios, predicted in the class of SUSY 
theories with see-saw mechanism 
of $\nu$-mass generation and soft 
SUSY breaking with universal boundary 
conditions at the GUT scale, on the 
Majorana CP-violation phases in the
PMNS neutrino mixing matrix 
% PMNS matrix  $\pmns$ 
and the  ``leptogenesis'' CP-violating (CPV) parameters.
The case of quasi-degenerate in mass heavy 
Majorana neutrinos is considered.
The analysis is performed 
for normal hierarchical 
(NH), inverted hierarchical (IH)
and quasi-degenerate (QD) light neutrino
mass spectra. We show, in particular, 
that for NH and IH $\nu$-mass spectrum 
and negligible lightest neutrino mass,
all three $l_i \to l_j + \gamma$
decay branching ratios,
% $BR(\mu \to e + \gamma)$, etc.,
$BR(l_i \to l_j + \gamma)$,
depend on one Majorana phase, 
one  leptogenesis CPV parameter
and on the 3-neutrino
oscillation parameters;
% $\theta_{12}$, $\theta_{23}$, $\theta_{13}$,
% $\delta$, $\deltasol$ and $\deltaatm$.
if the CHOOZ mixing angle $\theta_{13}$ is
sufficiently large, they depend on the 
Dirac CPV phase in the PMNS matrix.
The ``double ratios'' 
$\text{R}(21/31) \sim BR(\mu \to e + \gamma)/BR(\tau \to e + \gamma)$ and
$\text{R}(21/32) \sim BR(\mu \to e + \gamma)/BR(\tau \to \mu + \gamma)$
are determined by these parameters.
The same Majorana phase enters 
into the NH and IH expressions for the 
effective Majorana mass in 
neutrinoless double beta decay, $\mefff$. 

\end{abstract}

\newpage
\section{\label{sec:intro}Introduction}

% \vspace{-0.4cm}
\hskip 1.0truecm 
The experiments with solar, 
atmospheric, reactor and accelerator 
neutrinos~\cite{sol,SKsolaratm,SNO123,KamLAND,K2K}
have provided during the last several years 
compelling evidence for the existence 
of non-trivial % 3-$\nu$ 
3-neutrino
mixing in the weak charged-lepton current 
(see, e.g.,~\cite{STPNu04}):
%%%%%%%%%%%%%%%%%%%
\begin{equation}
\nu_{l \mathrm{L}}  = \sum_{j=1}^{3} U_{l j} \, \nu_{j \mathrm{L}},~~
l  = e,\mu,\tau,
\label{3numix}
\end{equation}
%%%%%%%%%%%%%%%%%%%
%
\noindent where 
$\nu_{lL}$ are the flavour neutrino fields, $\nu_{j \mathrm{L}}$ is
the field of neutrino $\nu_j$ having a mass $m_j$ and $U$ is the
Pontecorvo--Maki--Nakagawa--Sakata (PMNS) mixing 
matrix~\cite{BPont57}, $U \equiv \pmns$.  
The existing data, including the data from the 
\hbeta experiments~\cite{MoscowH3Mainz} 
imply that the massive neutrinos $\nu_j$ 
are significantly lighter than the
charged leptons and quarks: $m_{j} < 2.3$ eV 
(95\% C.L.)~\footnote{More stringent upper limit on $m_j$ follows from the
  constraints on the sum of neutrino masses obtained from
  cosmological/astrophysical observations, namely, the CMB data of the
  WMAP experiment combined with data from large scale structure
  surveys (2dFGRS, SDSS)~\cite{WMAPnu}: $\sum_{j} m_j < (0.7 - 2.0)$
  eV (95\% C.L.), where we have included a conservative estimate of
  the uncertainty in the upper limit (see, $\eg$,~\cite{Hanne03}).  }.

   The existence of the flavour neutrino mixing, eq.~(\ref{3numix}),
implies that the individual lepton charges, $L_l$, $l =e,\mu,\tau$,
are not conserved (see, $\eg$,~\cite{BiPet87}),
and processes like $\mu^- \rightarrow e^- + \gamma$,
$\mu^{-} \rightarrow e^{-} + e^{+} + e^{-}$, $\tau^- \rightarrow e^- +
\gamma$, $\tau^- \rightarrow \mu^- + \gamma$, $\mu^{-} + (A,Z)
\rightarrow e^{-} + (A,Z)$, $\etc$.  should take place.
Stringent experimental upper limits on
the branching ratios and relative cross-sections
of the indicated $|\Delta L_l| = 1$ decays and reactions
have been obtained
~\cite{mega,PDG04,BaBar05} (90\% C.L.):
%%%%%%%%%%%%%%%%%%%%%%%%%%%%%%%%%%%%%%%
\beq
\ba
\text{BR}(\mu \ra e+\gamma) < 1.2\times 10^{-11},~~~%\\[0.24cm]
\text{BR}(\mu \ra 3e) < 1.2\times 10^{-12}~,~\\[0.24cm]
\text{BR}(\tau \ra \mu +\gamma) < 6.8\times 10^{-8}~,~~~%\\[0.24cm]
\text{R}(\mu^{-} + \text{Ti} \rightarrow e^{-} + \text{Ti}) <
4.3\times 10^{-12}.  \ea \eeq
%%%%%%%%%%%%%%%%%%%%%%%%%%%%%%%%%%%%%%%%
%
%\vspace{-0.2cm}
\noindent Future experiments with increased 
sensitivity can reduce the current bounds on $\text{BR}(\mu\ra
e+\gamma)$, $\text{BR}(\tau\ra \mu +\gamma)$ and on $\text{R}(\mu^{-}
+ (A,Z) \rightarrow e^{-} + (A,Z))$ by a few orders of magnitude (see,
$\eg$,~\cite{Kuno99}).  In the experiment MEG under preparation at 
PSI~\cite{psi} it is planned to reach a sensitivity to
%%%%%%%%%%%%%%%%%%%%%%%%%%%%%
\begin{equation}
 \text{BR}(\mu \ra e+\gamma)\sim (10^{-13} - 10^{-14})\,. 
\end{equation}
%%%%%%%%%%%%%%%%%%%%%%%%%%%%%
%
\indent In the minimal extension of the Standard Theory
with massive neutrinos and neutrino mixing,
the rates and cross sections of the 
LFV processes are suppressed by the factor~\cite{SP76} 
(see also~\cite{BPP77}) $(m_{j}/M_W)^4 < 6.7\times 10^{-43}$, $M_W$ being the
$W^{\pm}$ mass, which  renders them unobservable.
It was shown in ~\cite{BorzMas86} that
in SUSY theories with see-saw mechanism of neutrino 
mass generation
\footnote{An integral part of the see-saw mechanism are the 
right-handed (heavy) Majorana neutrinos \cite{Pont67}.}
\cite{seesaw} and soft SUSY breaking 
with universal boundary conditions at a scale $M_X$
above the right-handed (RH) Majorana neutrino mass scale $M_R$, $M_{X}>M_R$ 
the rates and cross sections of the LFV processes can be strongly 
enhanced and can be within the sensitivity 
of presently operating and future planned experiments
(see also, e.g.,
\cite{Hisano96,Iba01,JohnE,Saclay0105,PPY03,PPR3,PPTY03,Eiichi05,PShinYasu05,SPWRTSYasu05}).
As is well-known,  the see-saw mechanism 
of neutrino mass generation~\cite{seesaw},
provides a very attractive explanation of the
smallness of the neutrino masses 
and - through the leptogenesis theory 
\cite{LeptoG}, of the observed baryon asymmetry 
of the Universe.

   One of the basic ingredients of the see-saw mechanism is the matrix of
neutrino Yukawa couplings, $\mathbf{Y_{\nu}}$.  Leptogenesis depends
on $\mathbf{Y_{\nu}}$ as well. In the large class of SUSY models with
see-saw mechanism and SUSY breaking mediated by flavour-universal 
soft terms at a scale $M_{X}>M_R$ we will consider,
the probabilities of LFV processes also depend strongly on
$\mathbf{Y_{\nu}}$ (see, $\eg$,~\cite{Iba01,JohnE}).  
The matrix $\mathbf{Y_{\nu}}$ can be expressed 
in terms of the light neutrino and heavy RH neutrino 
masses, the neutrino mixing matrix $\pmns$, and an 
orthogonal matrix $\mathbf{R}$~\cite{Iba01}.
Obviously, $\mathbf{Y_{\nu}}$ 
depends on the Majorana CP-violation
(CPV) phases in the PMNS matrix $\pmns$ ~\cite{BHP80}. 
In the case of negligible flavour effects in leptogenesis
\cite{flavorLG1}, 
successful leptogenesis is possible only if 
$\mathbf{R}$ is complex (see, e.g., \cite{LeptoG1}). 
For $M_R \ltap 10^{12}$ GeV, 
the flavour effects in leptogenesis 
can be  substantial \cite{flavorLG1} 
(see also \cite{flavorLG2,flavorLG3}). 
Due to the latter, leptogenesis can take place 
even for real $\mathbf{R}$ \cite{flavorLG1,PPRi106}, 
but only if $\mathbf{R} \neq {\bf 1}$. 
% Successful leptogenesis was shown to be
% possible for real $\mathbf{R} \neq {\bf 1}$, e.g.,
% for hierarchical and quasi-degenerate 
% light and heavy Majorana neutrino mass spectra 
% \cite{PPRi106}. 
In this case the Dirac and/or the Majorana 
CPV phases in the PMNS matrix play the role 
of the CPV parameters responsible for 
the generation of the baryon asymmetry 
of the Universe. Thus, in this case 
there is a direct link between the 
low-energy leptonic CP-violation and the 
generation of the baryon asymmetry of the 
Universe \cite{PPRi106}.

  It was shown in~\cite{PPY03} that if both 
the light and the heavy Majorana neutrino 
mass spectra are quasi-degenerate (QD),
the rates of LFV decays $\mu\rightarrow e + \gamma$, 
$\tau \rightarrow e + \gamma$ and
$\tau \rightarrow \mu + \gamma$,  
predicted in the class of SUSY theories 
of interest, can be strongly enhanced by 
the leptogenesis CP-violating (CPV) parameters
in the complex matrix $\mathbf{R}$,
with respect to the rates predicted for   
real $\mathbf{R} \neq \mathbf{1}$ or for
$\mathbf{R} = \mathbf{1}$.
The indicated LFV decay rates were also noticed
in \cite{PPY03} to depend for complex 
$\mathbf{R} \neq \mathbf{1}$
on the Majorana CPV phases in $\pmns$.
This dependence was investigated recently
in \cite{PShinYasu05}
by taking into account the effects of the 
phases in the renormalisation 
group (RG) running of the light 
$\nu$-masses $m_j$ and of the mixing angles in $\pmns$.
It was found \cite{PShinYasu05} that
the Majorana phases can affect significantly 
the predictions for the
$\mu\rightarrow e+\gamma$ and $\tau\rightarrow e+\gamma$ decay rates.

   In the present article we extend the analyses 
performed in \cite{PPY03,PShinYasu05} 
% (see also \cite{SPWRTSYasu05}) 
to the cases of normal hierarchical and 
inverted hierarchical light neutrino mass spectra. 
We investigate also in greater detail the case of QD spectrum. 
More specifically, working in the framework of 
the class of SUSY theories with see-saw
mechanism and soft SUSY breaking with flavour-universal 
% soft SUSY breaking 
boundary conditions at a scale $M_X>M_R$,
we study in detail the 
dependence of the rates of charged lepton flavour
violating (LFV) radiative decays $\mu\rightarrow e + \gamma$, 
$\tau \rightarrow e + \gamma$ and
$\tau \rightarrow \mu + \gamma$,
on the Majorana CPV phases in $\pmns$ 
and on the leptogenesis CPV parameters 
in the complex orthogonal matrix $\mathbf{R}$. 
The case of quasi-degenerate (QD) in mass heavy 
RH Majorana neutrinos is considered.
It is well-known that in the case of 
heavy Majorana neutrinos with QD mass spectrum
(i.e., negligible splitting between the masses),
the rates of LFV radiative decays
of interest do not depend on the 
matrix $\mathbf{R} \neq 1$ if $\mathbf{R}$ is 
a real matrix (see, e.g., \cite{PPY03}).
Our analysis is performed under the condition 
of negligible RG effects for the light neutrino 
masses $m_j$ and the mixing angles 
and CP-violation phases in $\pmns$.
The RG effects in question 
(see, e.g., \cite{RGrunU,PShinYasu05} 
and the references quoted therein)
are negligibly small
in the class of SUSY theories we are considering
in the case of hierarchical (normal or inverted) 
% light neutrino 
$\nu_j$ mass spectrum. 
The same is valid for QD
$\nu_j$ mass spectrum provided the SUSY 
parameter $\tan\beta$ is relatively small,
$\tan\beta < 10$, $\tan\beta$ being the ratio of the
vacuum expectation values of the up- and down- type
Higgs doublet fields in SUSY extensions of the 
Standard Theory. For the three types of 
light neutrino mass spectrum,
we investigate  also the predictions
for the ratios of the rates of  
$\mu\rightarrow e + \gamma$ and $\tau \rightarrow e + \gamma$,
and of $\mu\rightarrow e + \gamma$ and
$\tau \rightarrow \mu + \gamma$, decays.
In a large region of the relevant SUSY parameter space
these two ratios are independent of the SUSY parameters 
and are determined completely 
by the neutrino mixing angles,
Majorana and Dirac CPV phases, leptogenesis
CPV parameter(s) and, depending on the type of
the neutrino mass spectrum - hierarchical or quasi-degenerate,
by the neutrino mass squared
differences $\deltasol$ and $\deltaatm$ or 
the absolute neutrino mass. A study of the predictions 
for the two LFV decay rate ratios was performed recently in 
ref. \cite{GIsid05}. In \cite{GIsid05}, however,
only the case of zero leptogenesis CPV 
parameters (i.e., of real matrix $\mathbf{R}$) 
and of zero Majorana CPV 
phases in $\pmns$ was investigated. 
%
%%%%%%%%%%%%%%%%%%%%%%%%%%%%%%%%%%%%%%%%%%%%%%
%
\section{\large{Neutrino Mixing Parameters from Neutrino Oscillation Data}}
%
%%%%%%%%%%%%%%%%%%%%%%%%%%%%%%%%%%%%%%%%%%%%%
%
\hskip 1.0truecm
We will use the standard parametrisation of the
PMNS matrix $\pmns$ (see, $\eg$,~\cite{BPP1}): 
%%%%%%%%%%%%%%%%%%%%%%%%%%%%%%%%%%%
\bea 
\label{eq:Upara}
\pmns = \left( \bad 
 c_{12} c_{13} & s_{12} c_{13} & s_{13} e^{-i \delta} \\[0.2cm]   
 -s_{12} c_{23} - c_{12} s_{23} s_{13} e^{i \delta} 
 & c_{12} c_{23} - s_{12} s_{23} s_{13} e^{i \delta} & s_{23} c_{13} 
% e^{i \delta} 
\\[0.2cm] 
 s_{12} s_{23} - c_{12} c_{23} s_{13} e^{i \delta} & 
 - c_{12} s_{23} - s_{12} c_{23} s_{13} e^{i \delta} & c_{23} c_{13} 
% e^{i \delta} 
\\ 
     \ea   \right) 
{\rm diag}(1, e^{i \frac{\alpha}{2}}, e^{i \frac{\beta_M}{2}}) \, ,
%   {\rm diag}(1, e^{i \alpha}, e^{i \beta_M}) \, , 
\eea
%%%%%%%%%%%%%%%%%%%%%%%%%%%%%%%%%
%
\noindent where 
$c_{ij} = \cos\theta_{ij}$, $s_{ij} = \sin\theta_{ij}$, the angles
$\theta_{ij} = [0,\pi/2]$, $\delta = [0,2\pi]$ is the Dirac
CP-violating phase and $\alpha$ and $\beta_M$ are two Majorana
CP-violation phases~\cite{BHP80,SchValle80D81}. One can identify the
neutrino mass squared difference responsible for solar neutrino
oscillations, $\dms$, with $\Delta m^2_{21} \equiv m^2_2 - m^2_1$,
$\dms = \Delta m^2_{21} > 0$.  The neutrino mass squared difference
driving the dominant $\nu_{\mu} \rightarrow \nu_{\tau}$
($\bar{\nu}_{\mu} \rightarrow \bar{\nu}_{\tau}$) oscillations of
atmospheric $\nu_{\mu}$ ($\bar{\nu}_{\mu}$) is then given by
$|\dma|=|\Delta m^2_{31}|\cong |\Delta m^2_{32}| \gg \Delta m^2_{21}$.
The corresponding solar and atmospheric neutrino mixing angles,
$\theta_{\odot}$ and $\theta_{\rm A}$, coincide with $\theta_{12}$ and
$\theta_{23}$, respectively.  The angle $\theta_{13}$ is limited by
the data from the CHOOZ and Palo Verde experiments~\cite{CHOOZPV}.

 The existing neutrino oscillation data allow us to determine $\Delta
m^2_{21}$, $|\Delta m^2_{31}|$, $\sin^2\theta_{12}$ and
$\sin^22\theta_{23}$ with a relatively good precision and to obtain
rather stringent limits on $\sin^2\theta_{13}$ (see,
$\eg$,~\cite{BCGPRKL2,Schwatm05}). The best fit values and the 95\%
C.L. allowed ranges of $\Delta m^2_{21}$, $\sin^2\theta_{12}$,
$|\Delta m^2_{31}|$ and $\sin^22\theta_{23}$ read
~\footnote{The data imply, in particular, that 
maximal  solar neutrino mixing is ruled out at $\sim 6\sigma$; at 95\% C.L.\ 
  one finds $\cos 2\theta_\odot \geq 0.26$~\cite{BCGPRKL2}, which has
  important implications~\cite{PPSNO2bb}.}:
%%%%%%%%%%%%%%%%%%%%%%%%%%%%%%%%%%%
\beq
\label{bfvsol}
\ba
\deltasol = 8.0\times 10^{-5}~{\rm eV^2},~~
\sin^2\theta_{21} = 0.31~, \\[0.25cm]
\deltasol = (7.3 - 8.5) \times 10^{-5}~{\rm eV^2},~~
\sin^2 \theta_{12} = (0.26 - 0.36)~,
\ea
\eeq
%%%%%%%%%%%%%%%%%%%%%%%%%%%%%%%%%%%%
% \vspace{-0.5cm}
%%%%%%%%%%%%%%%%%%%%%%%%%%%%%%%%% 
\beq 
\label{eq:atmrange}
\ba
% |\dma| = 2.2\times 10^{-3}~{\rm eV^2}~,~~\sin^22\theta_{\rm A} = 1.0
|\deltaatm| =2.2\times 10^{-3}~{\rm eV^2}~,~~\sin^22\theta_{23} = 1.0
~, \\  [0.25cm]
% |\dma| = (1.3 - 4.2)\times 10^{-3}~{\rm eV^2},~~
% \sin^22\theta_{\rm A} \geq 0.85. SK  99.73\% C.L.
% |\dma| = (1.4 - 3.3)\times 10^{-3}~{\rm eV^2},~~
% \sin^22\theta_{\rm A} \geq 0.85. % Schwetz 99.73 \% C.L.
% |\dma| = (1.7 - 2.9)\times 10^{-3}~{\rm eV^2}~,~~
% 
%\sin^22\theta_{\rm A} \geq 0.90. % Schwetz 95\% C.L.
|\deltaatm| = (1.7 - 2.9)\times 10^{-3}~{\rm eV^2}~,~~
\sin^22\theta_{23} \geq 0.90. 
\ea
\eeq
%%%%%%%%%%%%%%%%%%%%%%%%%%%%%%%%%
% 
\noindent
A combined
\footnote{Using the recently
announced (but still unpublished) 
data from the MINOS experiment \cite{MINOS0306} 
in the analysis leads to somewhat different best fit
value and 95\% allowed range of $|\deltaatm|$
\cite{TSchw0406}:
$|\deltaatm| = 2.5\times 10^{-3}~{\rm eV^2}$
and $|\deltaatm| = (2.2 - 2.9)\times 10^{-3}~{\rm eV^2}$.}
3-$\nu$ oscillation analysis of the solar 
neutrino, KL and CHOOZ data gives~\cite{BCGPRKL2}
%%%%%%%%%%%%%%%%%%%%%%%%%%%%%%
\beq
\sin^2\theta_{13} < 0.027~(0.044),~~~~\mbox{at}~95\%~(99.73\%)~{\rm C.L.}
\label{th13}
\eeq
%%%%%%%%%%%%%%%%%%%%%%%%%%%%%%%
%
The neutrino oscillation parameters discussed above can (and very
likely will) be measured with much higher accuracy in the future (see,
$\eg$,~\cite{STPNu04}).

  The sign of $\dma = \deltaatm $, as it is well known, cannot be
determined from the present (SK atmospheric neutrino and K2K) data.
The two possibilities, $\Delta m^2_{31(32)} > 0$ or $\Delta
m^2_{31(32)} < 0$ correspond to two different
types of $\nu$-mass spectrum:\\
-- {\it with normal ordering (hierarchy)} 
$m_1 < m_2 < m_3$, $\dma=\Delta m^2_{31} >0$, and \\
-- {\it with inverted ordering (hierarchy)} 
$m_3 < m_1 < m_2$, $\dma =\Delta m^2_{32}< 0$. \\
\noindent Depending on the sign of \dma, ${\rm sgn}(\dma)$, and 
the value of the lightest neutrino mass,
${\rm min}(m_j)$, the $\nu$-mass  spectrum can be\\
-- {\it Normal Hierarchical}: $m_1{\small \ll m_2 \ll }m_3$,
$m_2{\small \cong }(\dmsol)^ {1\over{2}}{\small \sim}$ 0.009 eV,
$m_3{\small \cong }|\dma|^{1\over{2}}{\small \sim}$ 0.05 eV;\\
-- {\it Inverted Hierarchical}: $m_3 \ll m_1 < m_2$,
with $m_{1,2} \cong |\dma|^{1\over{2}}\sim$ 0.05 eV; \\
-- {\it Quasi-Degenerate (QD)}: $m_1 \cong m_2 \cong m_3 \cong m$,
$m_j^2 \gg |\dma|$, $m \gtap 0.10$~eV.

 The sign of $\Delta m^2_{31} \cong \Delta m^2_{32}$, 
which drives the dominant atmospheric neutrino 
oscillations, can be 
determined by studying 
oscillations of neutrinos and
antineutrinos, say, 
$\nu_{\mu} \rightarrow \nu_e$
and $\bar{\nu}_{\mu} \rightarrow \bar{\nu}_e$,
in which matter effects are sufficiently large.
This can be done, e.g., in long-baseline 
$\nu$-oscillation experiments 
(see, e.g.,~\cite{AMMS99}).
Information about ${\rm sgn}(\Delta m^2_{31})$
can be obtained also in atmospheric neutrino 
experiments by studying 
the oscillations of the
atmospheric $\nu_{\mu}$ and $\bar{\nu}_{\mu}$ which
traverse the Earth \cite{JBSP203}.

  As is well-known, the theories employing 
the see-saw mechanism of neutrino mass 
generation~\cite{seesaw} of interest for our 
discussion, predict the massive neutrinos 
$\nu_j$ to be Majorana particles. Determining 
the nature of massive neutrinos is one of the
most formidable and pressing problems in today's neutrino physics
(see, $\eg$,~\cite{STPNu04,APSbb0nu}). 
If it is established that the massive neutrinos $\nu_j$ 
are indeed Majorana fermions, getting information about 
the Majorana CP-violation phases in $\pmns$, 
would be a very difficult problem.
The oscillations of flavour neutrinos, 
$\nu_{l} \rightarrow \nu_{l'}$ and 
$\bar{\nu}_{l} \rightarrow \bar{\nu}_{l'}$,
$l,l'=e,\mu,\tau$, are insensitive to the Majorana CP-violation phases
$\alpha$ and $\beta_M$~\cite{BHP80,Lang87}.
The only feasible experiments that at present have the potential of
establishing the Majorana nature of light neutrinos $\nu_j$ and of
providing information on the Majorana CP-violation phases in $\pmns$
are the experiments searching for the neutrinoless double beta
($\betabeta$-) decay, $(A,Z) \rightarrow (A,Z+2) + e^- + e^-$ (see,
$\eg$,~\cite{BiPet87,APSbb0nu,STPFocusNu04}). 
The $\betabeta$-decay effective Majorana mass, $\mefff$ 
(see, e.g., \cite{BiPet87}), which contains all the 
dependence of the $\betabeta$-decay amplitude on 
the neutrino mixing parameters, 
is given by the following expressions 
for the normal hierarchical (NH),  
inverted hierarchical (IH) 
and quasi-degenerate (QD) neutrino mass spectra
(see, e.g., \cite{STPFocusNu04}): 
%%%%%%%%%%%%%%%%%%%%%%%%%%%%%%%%%%%%%%%%%%%%
\begin{eqnarray}
\meff&\cong&\left|\sqrt{\deltasol}\sin^2 \theta_{12}e^{i\alpha}
+\sqrt{\deltaatm}\sin^2\theta_{13}
e^{i\beta_{M}} \right|\;,~m_1\ll m_2 \ll m_3~{\rm (NH)},
\label{meffNH2}
\end{eqnarray}
%%%%%%%%%%%%%%%%%%%%%%%%%%%%%%%
%%%%%%%%%%%%%%%%%%%%%%%%%%%%%%%
\begin{eqnarray}
\meff &\cong& \sqrt{\Delta m^2_{13}}
\left|\cos^2\theta_{12}  + 
e^{i\alpha}~\sin^2 \theta_{12} \right|\;,~~m_3 \ll m_1< m_2~{\rm (IH)}, 
\label{meffIH1}
\end{eqnarray}
%%%%%%%%%%%%%%%%%%%%%%%%%%%%%%%%
%%%%%%%%%%%%%%%%%%%%%%%%%%%%%%%%
\begin{eqnarray}
\meff &\cong& m \left|\cos^2\theta_{12}
 +  e^{i \alpha}~\sin^2 \theta_{12} 
% +   m_3\sin^2\theta_{13}~e^{i\alpha_{31}} 
 \right|\;,~~m_{1,2,3} \cong m \gtap 0.10~{\rm eV}\;~{\rm (QD)}. 
\label{meffQD0} 
\end{eqnarray}
%%%%%%%%%%%%%%%%%%%%%%%%% 
%
Obviously, $\meff$ depends strongly 
on the Majorana CP-violation phase(s)
\footnote{We assume that the fields of the
Majorana neutrinos
$\nu_j$ satisfy the Majorana conditions:
$C(\bar{\nu}_{1,2})^{T} = \nu_{1,2}$, and
$C(\bar{\nu}_{3})^{T} = e^{-i 2\delta} \nu_{3}$,
where $C$ is the charge conjugation matrix.
With the parametrisation we are employing for
$\pmns$, eq. (\ref{eq:Upara}), the effective Majorana 
mass $\meff$ does not depend on the Dirac CP-violation
phase $\delta$ as a consequence of
the presence of the phase factor
$e^{-i 2\delta}$ in the Majorana condition
for the field $\nu_{3}$.};
% Given $\deltasol$, $|\deltaatm|$, $\theta_{12}$,
% and $\theta_{13}$ ($|\deltaatm|$ ($m$) and $\theta_{12}$),
% % given $|\deltaatm|$ ($m_0$) and $\theta_{12}$,
the CP-conserving values of 
$(\alpha - \beta_M) = 0,\pm \pi$
($\alpha =0,\pm\pi$) \cite{LW81}, 
in particular, determine the range of 
possible values of $\meff$ in the 
case of NH (IH, QD) spectrum.
If the $\betabeta$-decay is observed, 
the measurement of the $\betabeta$-decay half-life
combined with information on the absolute scale of neutrino masses (or
on ${\min}(m_j)$), might allow to significantly 
constrain the Majorana phase
$\alpha$ \cite{BPP1,BGKP96,PPW}, for instance. 
%
%%%%%%%%%%%%%%%%%%%%%%%%%%%%%%%%%%%%%%
%
\section{\large{The See-Saw Mechanism, Neutrino Yukawa Couplings, 
and LFV Decays $l_i \to l_j +\gamma$}}
%
%%%%%%%%%%%%%%%%%%%%%%%%%%%%%%%%%%%%%%%%%%
% 
\hskip 1.0truecm In the 
minimal supersymmetric standard model
with RH neutrinos $N_j$ and see-saw mechanism of 
neutrino mass generation (MSSMRN) we consider
it is always possible to choose a basis in which both
the matrix of charged lepton Yukawa couplings, 
$\mathbf{Y_{\rm E}}$, and the Majorana mass matrix of the 
heavy RH neutrinos, $\mathbf{M_{\rm N}}$, are real and diagonal.
We will work in that basis and will denote by 
$\mathbf{D_{\rm N}}$ the corresponding diagonal 
RH neutrino mass matrix, $\mathbf{D_{\rm N}} =
{\rm diag}(M_1,M_2,M_3)$, with $M_j > 0$. 
We will consider in what follows the case of
QD heavy Majorana neutrinos:
$M_1 \cong M_2 \cong M_3 = M_R$.
It will be assumed that 
the splittings between the masses of the heavy 
Majorana neutrinos are sufficiently small, e.g.,
that they are of the order of those considered 
in \cite{PPY03}. The existence of 
sufficiently small (but nonzero) 
splittings between the masses 
of the heavy Majorana neutrinos $N_j$ is indeed a 
necessary condition for the successful
(resonant) leptogenesis to take place.
The requisite small mass splittings
can be generated, e.g., by renormalisation group
effects \cite{GonzF03}. 
However, the mass splittings under discussion,
$|M_i - M_j| \ll M_i,M_j$, $i\neq j =1,2,3$,  
do not play any significant role in the 
predictions for the rates of the
decays $l_i \rightarrow l_j + \gamma$,
which is the main subject of the present study.
The heavy Majorana neutrino mass $M_R$ 
will standardly be assumed to be 
smaller than the GUT scale 
$M_{\rm GUT} \simeq 2\times 10^{16}$ GeV.

 In the class of theories of interest,
the branching ratio of the $l_i\to l_j + \gamma$ decay 
has the following form (in the ``mass insertion'' 
and leading-log approximations,
see, e.g.,~\cite{Hisano96,JohnE,PPTY03}):
%%%%%%%%%%%%%%%%%%%%%%%%%%%%%%%%%
\begin{align}
\text{BR}(l_i\to l_j\gamma)\cong &
% \frac{\Gamma(l_i\to e\nu\bar{\nu})}{\Gamma_{\text{total}}(l_i)}
% \frac{\alpha_{\text{em}}^3}{G_F^2}
% \frac{|(M_L^2)_{ij}|^2}{m_S^8}\tan^2\beta\nonumber\\
% \simeq&
\frac{\Gamma(l_i\to e\nu\bar{\nu})}{\Gamma_{\text{total}}(l_i)}
\frac{\alpha_{\text{em}}^3}{G_F^2m_S^8}
\left|\frac{(3 + a_0^2)m_0^2}{8\pi^2}\right|^2
\left|\sum_k (Y_{\nu}^{\dagger})^{ik}~\ln\frac{M_{X}}{M_k}~Y_{\nu}^{kj}
\right|^2\tan^2\beta\;,
\label{eq_ijg}
\end{align}
%%%%%%%%%%%%%%%%%%%%%%%%%%%%%%%%%%%%%%%%%%
%
where $i\neq j=1,2,3$, $l_1,l_2,l_3\equiv e,\mu,\tau$,
$m_0$ and $A_0 = a_0m_0$ are the universal SUSY breaking
scalar masses and trilinear scalar couplings at 
$M_X > M_R$, $m_S$ represents SUSY particle mass (see further),
$\tan\beta$ is the ratio of the 
vacuum expectation values of up-type and down-type
Higgs fields and $\mathbf{Y}_{\nu} = \mathbf{Y}_{\nu}(M_R)$ 
is the matrix of neutrino Yukawa couplings evaluated at $M_R$.
The matrix $\mathbf{Y}_{\nu}$ can be parametrised  as~\cite{Iba01}
%%%%%%%%%%%%%%%%%%%%%%%%%%%%
\begin{align}
  \mathbf{Y}_{\nu}(M_R) = \frac{1}{v_u}
  \sqrt{\mathbf{D}_N}~\mathbf{R}~
  \sqrt{\mathbf{D}_{\nu}}~\mathbf{U}^{\dagger}
\cong 
\frac{1}{v_u} \sqrt{M_R}~\mathbf{R}~
  \sqrt{\mathbf{D}_{\nu}}~\mathbf{U}^{\dagger} \;.
%  \mathbf{Y}_{\nu}(M_R) = \frac{1}{v_u}
%  \sqrt{\mathbf{D}_N}(M_R)~\mathbf{R}~
%  \sqrt{\mathbf{D}_{\nu}}(M_R)~\mathbf{U}^{\dagger}(M_R)\;.
\label{eq_para_yn}
\end{align}
%%%%%%%%%%%%%%%%%%%%%%%%%%%
%
Here $v_u = v \sin\beta$, where $v = 174$ GeV,
$\mathbf{R}$ is a complex orthogonal matrix 
\footnote{Equation (\ref{eq_para_yn}) represents the so-called
``orthogonal'' parametrisation of  $\mathbf{Y}_{\nu}$.
In certain cases it is more convenient to use the
``bi-unitary'' parametrisation \cite{PPR3}
$\mathbf{Y}_{\nu} = 
\mathbf{U}^{\dagger}_{R} \mathbf{Y}^{\rm diag}_{\nu}~\mathbf{U}_{L}$,
where $\mathbf{U}_{\rm L,R}$ are unitary matrices and
$\mathbf{Y}^{\rm diag}_{\nu}$ is a real diagonal matrix.
The orthogonal parametrisation is better adapted
for our analysis and we will employ it in what follows.}
$\mathbf{R}^T\mathbf{R}= \mathbf{1}$,
$\mathbf{D}_{\nu} = \mathrm{diag}(m_1,m_2,m_3)$,
%$m_1\geq 0$ and $m_{2,3} > 0$ being the light neutrino masses,
$m_{1,2,3} > 0$ being the light neutrino masses
\footnote{To be more precise, 
we can have ${\rm min}(m_j)=0$.} 
and $\mathbf{U}$ is the PMNS matrix. 

  In what follows we will consider the case when
the RG running of $m_j$ and of the parameters in
$\pmns$ from approximately $M_Z \sim 100$ GeV,
where they are measured, to $M_R$ is relatively small 
and can be neglected. This possibility is realised 
in the class of theories under discussion for 
sufficiently small values of $\tan\beta$ 
and/or of the lightest neutrino mass ${\rm min}(m_j)$,
e.g., for $\tan\beta < 10$ and/or ${\rm min}(m_j) \ltap 0.05$ eV 
(see, e.g., \cite{RGrunU,PShinYasu05}).
Under the indicated condition, 
% we have
%%%%%%%%%%%%%%%%%%%%%%%%%%%%
% \begin{align}
%  \mathbf{Y}_{\nu}(M_R) \cong \frac{1}{v_u}
%  \sqrt{\mathbf{D}_N}(M_R)~\mathbf{R}~
%  \sqrt\mathbf{D}_{\nu}{}(M_Z)~\mathbf{U}^{\dagger}(M_Z)\;.
% \label{eq_para_yn}
% \end{align}
%%%%%%%%%%%%%%%%%%%%%%%%%%%
%
$\mathbf{D}_{\nu}$ and $\mathbf{U}$
in eq. (\ref{eq_para_yn}) should be taken 
at the scale $\sim M_Z$, at which the neutrino 
mixing parameters are measured.

  It was shown in~\cite{PPTY03} that in 
a large region of the relevant soft 
SUSY breaking parameter space, the expression
%%%%%%%%%%%%%%%%%%%%%%%%%%%%%%%%%%%%%%%%%%
\begin{align}
m_S^8\simeq 0.5~m_0^2~m_{1/2}^2~(m_0^2 + 0.6 ~m_{1/2}^2)^2\;,
\label{eq_ms}
\end{align}
%%%%%%%%%%%%%%%%%%%%%%%%%%%%%%%%%%%%%%%%%%
%
$m_{1/2}$ being the universal gaugino 
mass at $M_X$, gives an excellent approximation 
to the results obtained in a full
renormalisation group analysis, i.e., 
without using the leading-log and 
the mass insertion approximations. 
For  values of the soft SUSY breaking parameters 
implying SUSY particle masses in the range of 
few to several hundred GeV, say,
$m_0 = m_{1/2} = 250$ GeV, 
$A_0 = a_0m_0 = -100$ GeV, we have:
%%%%%%%%%%%%%%%%%%%%%%%%%%%%%%%%%%%%%%%%%%
\be \label{eq:bench}
% BR(l_i \ra l_j \gamma) \cong 2.0 \cdot 10^{-9}
BR(l_i \ra l_j \gamma) \cong 9.1 \times 10^{-10} 
\left| (\mathbf{Y_{\nu}^\dagger} L \mathbf{Y_{\nu}})_{ij} \right|^2 
\, \tan^2 \beta ~,
\ee
%%%%%%%%%%%%%%%%%%%%%%%%%%%%%%%%%%%%%%%%%%%
%
where $L \cong \ln(M_{X}/M_R)$.
Since $\tan^2 \beta \gtap 10$,
eq. (\ref{eq:bench}) implies that if indeed the SUSY particle masses
do not exceed several hundred GeV, 
the quantity $|(\mathbf{Y_{\nu}^\dagger} L \mathbf{Y_{\nu}})_{21}|$
has to be relatively small. 
This is realised for, e.g., $M_R \ltap 10^{12}$ GeV.

   As follows from eqs. (\ref{eq_ijg}) and (\ref{eq:bench})
and was widely discussed , in the case of soft SUSY
breaking mediated by soft 
flavour-universal terms at $M_X>M_R$, the
predicted rates of LFV processes 
such as $\mu\to e + \gamma$ decay are
very sensitive to the off-diagonal elements of
%%%%%%%%%%%%%%%%%%%%%%%%%%%%%%%%%%%%%
\begin{align}
\mathbf{Y}_{\nu}^{\dagger}(M_R)\mathbf{Y}_{\nu}(M_R)
= \frac{1}{v_u^2}~
\mathbf{U}\sqrt{\mathbf{D}_{\nu}}~\mathbf{R}^{\dagger}~
\mathbf{D}_N~\mathbf{R}~\sqrt{\mathbf{D}_{\nu}}
\mathbf{U}^{\dagger}
\cong 
\frac{M_R}{v_u^2}~
\mathbf{U}\sqrt{\mathbf{D}_{\nu}}~\mathbf{R}^{\dagger}
~\mathbf{R}~\sqrt{\mathbf{D}_{\nu}}
\mathbf{U}^{\dagger}
\;.
\label{YnudYnu}
\end{align}
%%%%%%%%%%%%%%%%%%%%%%%%%%%%%%%%%%%%%
%

 It is well-known that in the theories with 
see-saw mechanism, leptogenesis depends 
on $\mathbf{Y}_{\nu}(M_R)$ and thus on $\mathbf{R}$. 
In the case of negligible flavour 
effects \cite{flavorLG1},
the dependence of interest is realised
through the product \cite{LeptoG1}
%%%%%%%%%%%%%%%%%%%%%%%%%%%%%%%%%%%%%
\begin{align}
\mathbf{Y}_{\nu}(M_R)\mathbf{Y}_{\nu}^{\dagger}(M_R)
= \frac{1}{v_u^2}~
\sqrt{\mathbf{D}_{\rm N}}~\mathbf{R}~
\mathbf{D}_{\nu}~\mathbf{R}^{\dagger}~\sqrt{\mathbf{D}_{\rm N}}
\cong 
\frac{M_R}{v_u^2}~\mathbf{R}~
\mathbf{D}_{\nu}~\mathbf{R}^{\dagger}\;.
\label{YnuYnud}
\end{align}
%%%%%%%%%%%%%%%%%%%%%%%%%%%%%%%%%%%%%
%
In this case successful leptogenesis 
can take place only if $\mathbf{R} \neq {\bf 1}$ 
is complex. If $M_R \ltap 10^{12}$ GeV,
flavour effects in leptogenesis can be significant
and leptogenesis can proceed successfully 
even for real  $\mathbf{R} \neq {\bf 1}$
(see, e.g., \cite{PPRi106}). It follows from eqs.
(\ref{eq_ijg}) and (\ref{YnudYnu}), however, 
that in the case of QD in mass heavy RH Majorana 
neutrinos of interest, the predicted rates of LFV decays 
$\mu\to e + \gamma$, etc. are independent of the 
orthogonal matrix $\mathbf{R}$ if $\mathbf{R}$ is real.

   In what follows we will consider 
$(\mathbf{R})^{*}\neq \mathbf{R}$ 
and  will use the parameterizations 
of $\mathbf{R}$ proposed in \cite{PPY03}:
%%%%%%%%%%%%%%%%%%%%%%%%%%%%%%%%%%%%%
\begin{align}
\mathbf{R} = \mathbf{O}~e^{i\mathbf{A}}\;.
\label{RPPY}
\end{align}
%%%%%%%%%%%%%%%%%%%%%%%%%%%%%%%%%%%%%
%
Here $\mathbf{O}$ is a {\it real orthogonal} 
matrix 
%%%%%%%%%%%%%%%%%%%%%%%%%%%%%%%%%%%%%
% \begin{align}
% \mathbf{O} = \mathbf{O}_{12} \, \mathbf{O}_{13} \, \mathbf{O}_{23}~, 
% \label{O3rot}
% \end{align}
%%%%%%%%%%%%%%%%%%%%%%%%%%%%%%%%%%%%%
%
% where $\mathbf{O}_{ij}$ are $2\times 2$ orthogonal 
% matrices of rotations with real angles $\rho_{ij}$,
and $\mathbf{A}$ is a {\it real antisymmetric} matrix, 
$(\mathbf{A})^T = - \mathbf{A}$,
%%%%%%%%%%%%%%%%%%%%%%%%%%%%%%%%%%%%%%%
\begin{align}
A=
\begin{pmatrix}
0&a&b\\
-a&0&c\\
-b&-c&0
\end{pmatrix}\;,
\label{Aabc}
\end{align}
%%%%%%%%%%%%%%%%%%%%%%%%%%%%%%%%%%%%
%
$a,b,c$ being real parameters. The following 
representation of $e^{i\mathbf{A}}$ 
proves useful ~\cite{PPY03}:
%%%%%%%%%%%%%%%%%%%%%%%%%%%%%%%%%%%%%%%
\begin{align}
% \mathbf{R} = 
e^{i\mathbf{A}} = 
\mathbf{1}-\frac{\cosh r-1}{r^2}\mathbf{A}^2+i\frac{\sinh r}{r}\mathbf{A}\,,
\label{exp}
\end{align}
%%%%%%%%%%%%%%%%%%%%%%%%%%
%
with $r=\sqrt{a^2+b^2+c^2}$. 
The requirement of successful
leptogenesis in the case of QD light and 
heavy Majorana neutrino mass
spectra and negligible flavour 
effects \cite{flavorLG1} implies~\cite{PPY03} 
that $abc \neq 0$ and that 
none of the parameters $|a|$, $|b|$ and $|c|$
can be exceedingly small: 
$|abc| \sim (10^{-6} - 10^{-4})$.  
One also finds from the condition that 
Yukawa couplings should have 
moduli which do not exceed $\sim 1$
that typically $r\ltap 1$  \cite{PPY03}.

  The parametrisation given in eq. (\ref{RPPY})
is particularly convenient in the analysis 
of the case of QD heavy Majorana neutrinos.
We will consider
a range of values of the parameters $a,b,c$
determined by $10^{-4} \ltap |a|,|b|,|c| \ltap 0.10$. 
Equations (\ref{eq_ijg}) and (\ref{YnudYnu}) imply that
for QD heavy Majorana neutrinos 
we can set $\mathbf{O} = \mathbf{1}$ 
and  use $\mathbf{R} = e^{i\mathbf{A}}$
in the calculation of $\text{BR}(l_i\to l_j+\gamma)$
without loss of generality. Results for
$\text{BR}(l_i\to l_j+\gamma)$ in the case of 
real $\mathbf{R} \neq {\bf 1}$ can be obtained 
by formally replacing $i\mathbf{A}$ by $\mathbf{0}$
in the expressions for  
$\text{BR}(l_i\to l_j+\gamma)$ derived 
using $\mathbf{R} = e^{i\mathbf{A}}$.

%%%%%%%%%%%%%%%%%%%%%%%%%%%%%%%%%%%%%%%%%%%%
%
\section{
\label{sec:main1}
\large{The LFV Decays $l_i\to l_j + \gamma$ and Majorana Phases}}
%
%%%%%%%%%%%%%%%%%%%%%%%%%%%%%%%%%%%%%%%%%%%%%%%%%%%
%
\hskip 1.0truecm In the case under discussion
$M_1 = M_2 = M_3 \equiv M_R$ and
the matrix of neutrino Yukawa couplings  
has the form $\mathbf{Y}_{\nu}=\frac{\sqrt{M_R}}{v_u}
\mathbf{O}e^{i\mathbf{A}}\sqrt{\mathbf{D}_{\nu}}\mathbf{U}^{\dagger}$.  The
off-diagonal elements of $\mathbf{Y}_{\nu}^{\dagger}\mathbf{Y}_{\nu}$
of interest do not depend on $\mathbf{O}$ and satisfy 
$(\mathbf{Y}_{\nu}^{\dagger}\mathbf{Y}_{\nu})^{*}_{ij}
= (\mathbf{Y}_{\nu}^{\dagger}\mathbf{Y}_{\nu})_{ji}$.
To leading order in small quantities they are given by
(see also \cite{PPY03,PShinYasu05})
%%%%%%%%%%%%%%%%%%%%%%%%%%%%%%%%%%%%
\begin{align}
\label{12R}
(Y_{\nu}^{\dagger}Y_{\nu})_{12}&= \Delta_{21}c_{23}c_{12}s_{12}
+\Delta_{31}s_{23}s_{13}e^{-i\delta}~~~~~~~~~~~~~~~~~~~~~~~~~~~~~~~~~~~~~~~~~~~~~~~~~~~~~~~~~~~~~~
%\Delta_{21}c_{23}c_{12}s_{12}c_{13} +\Delta_{31}s_{23}c_{13}s_{13}e^{-i\delta}
\nonumber\\
&+2\frac{M_R}{v_u^2} i\biggl[ a\sqrt{m_1m_2} \left (
  c_{23}(c_{12}^2e^{-i\frac{\alpha}{2}}+
  s_{12}^2e^{i\frac{\alpha}{2}})
% \right.
% \nonumber\\
% &\phantom{Space}
% \left.
% + s_{13}c_{12}s_{12}s_{23}(e^{i(\frac{\alpha}{2}-\delta)} - 
% e^{-i(\frac{\alpha}{2} +\delta)}) \right )
+ 2i~s_{13}c_{12}s_{12}s_{23}~e^{-i\delta}\sin\frac{\alpha}{2} \right )
\nonumber\\
&\phantom{Space}
+b\sqrt{m_1m_3}s_{23} \left ( c_{12}e^{-i\frac{\beta_M}{2}}
 -s_{13}s_{12}e^{i(\frac{\beta_M}{2}-\delta)} \right )
\nonumber\\
&\phantom{Space}
+ c\sqrt{m_2m_3} \left(s_{23}s_{12}e^{i\frac{\alpha-\beta_M}{2}}
-c_{23} s_{13}c_{12}e^{-i(\frac{\alpha-\beta_M}{2} + \delta)} \right )
+\mathcal{O}(s^2_{13})
\biggr]+\mathcal{O}(r^2,s_{13}^2)\;,%\\ % \nonumber\\
\end{align}
%%%%%%%%%%%%%%%%%%%%%%%%%%%%%%%
\begin{align}
\label{13R}
(Y_{\nu}^{\dagger}Y_{\nu})_{13}&= -\Delta_{21}s_{23}c_{12}s_{12}
+\Delta_{31}c_{23}s_{13}e^{-i\delta}~~~~~~~~~~~~~~~~~~~~~~~~~~~~~~~~~~~~~~~~~~~~~~~~~~~~~~~~~~~~~~
%-\Delta_{21}s_{23}c_{12}s_{12}c_{13}+\Delta_{31}c_{23}c_{13}s_{13}e^{-i\delta}
\nonumber\\
&+2
\frac{M_R}{v_u^2} i\biggl[
a\sqrt{m_1m_2} \left ( -s_{23}(c_{12}^2e^{-i\frac{\alpha}{2}}+
s_{12}^2e^{i\frac{\alpha}{2}}) 
% + s_{13}c_{12}s_{12}c_{23}(e^{i(\frac{\alpha}{2}-\delta)} - 
% e^{-i(\frac{\alpha}{2} +\delta)})
+ 2i~s_{13}c_{12}s_{12}c_{23}~e^{-i\delta}\sin\frac{\alpha}{2}
\right )
\nonumber\\
&\phantom{Space}
+b\sqrt{m_1m_3} \left ( c_{12}c_{23}e^{-i\frac{\beta_M}{2}}
 -s_{13}s_{12}s_{23}e^{i(\frac{\beta_M}{2}-\delta)} \right )
\nonumber\\
&\phantom{Space}
+ c\sqrt{m_2m_3}\left(s_{12}c_{23}e^{i\frac{\alpha-\beta_M}{2}}
+ s_{13}c_{12}s_{23} e^{-i(\frac{\alpha-\beta_M}{2} + \delta)} \right )
+\mathcal{O}(s^2_{13})
\biggr]+\mathcal{O}(r^2,s_{13}^2)\;,% \\ % \nonumber\\
\end{align}
%%%%%%%%%%%%%%%%%%%%%%%%%%%%%%%
\begin{align}
\label{23R}
(Y_{\nu}^{\dagger}Y_{\nu})_{23}&= \Delta_{31}s_{23}c_{23}~~~~~~~~~~~~~~~~~~~~~~~~~~~~~~~~~~~~~~~~~~~~~~~~~~~~~~~~~~~~~~~~~~~~~~~~~
% \Delta_{31}s_{23}c_{23}c_{13}^2
\nonumber\\
&+2\frac{M_R}{v_u^2}
i\biggl[
a\sqrt{m_1m_2} \left (-2i c_{12}s_{12}c_{23}s_{23}\sin\frac{\alpha}{2}
+ s_{13} \left [ c_{12}^2(s_{23}^2e^{-i(\frac{\alpha}{2}-\delta)} 
+ c_{23}^2e^{i(\frac{\alpha}{2} -\delta)}) \right. \right.
\nonumber\\
&\phantom{Space}
\left. \left.
+ s_{12}^2(s_{23}^2e^{i(\frac{\alpha}{2} +\delta)} 
+ c_{23}^2e^{-i(\frac{\alpha}{2} +\delta)})\right ]
\right )
\nonumber\\
&\phantom{Space}
+b\sqrt{m_1m_3} \left (-s_{12}(s_{23}^2e^{i\frac{\beta_M}{2}}
+c_{23}^2e^{-i\frac{\beta_M}{2}})
+s_{13}c_{12}c_{23}s_{23}~2i\sin(\frac{\beta_M}{2}-\delta)
\right )
\nonumber\\
&\phantom{Space}
+ c\sqrt{m_2m_3}\left ( c_{12}
(c_{23}^2e^{i\frac{\alpha-\beta_M}{2}}+
s_{23}^2e^{-i\frac{\alpha-\beta_M}{2}})
-s_{13}s_{12}c_{23}s_{23}~2i\sin(\frac{\alpha-\beta_M}{2} + \delta) 
\right )
\nonumber\\
&\phantom{Space}
+\mathcal{O}(s^2_{13}) \biggr]+\mathcal{O}(r^2,s_{13}^2)\;,
\end{align}
%%%%%%%%%%%%%%%%%%%%%%%%%%%%%%%%%%%%%%%%%%%%%
%
where 
%%%%%%%%%%%%%%%%%%%%%%%%%%%%%%5
\begin{align}
\Delta_{ij} \equiv
\frac{M_R}{v_u^2} \left(m_i -m_j \right) = 
\frac{M_R}{v_u^2}~\frac{\Delta m^2_{ij}}{m_i + m_j}\;.
\label{Dij}
\end{align}
%%%%%%%%%%%%%%%%%%%%%%%%%%%%%%%%%%%%%%%%%%%%%%%
%
Equations (\ref{12R})-(\ref{23R})
are valid for any of the possible 
types of light neutrino mass spectrum.
The corresponding expressions 
for real $\mathbf{R} \neq {\bf 1}$
can be obtained by formally 
setting the three leptogenesis CPV parameters
$a$, $b$ and $c$ to 0 in eqs. (\ref{12R})-(\ref{23R}).

 In what follows we will concentrate 
on the case of complex $\mathbf{R} \neq {\bf 1}$
and will call the real quantities
$a,b,c$ ``leptogenesis CP-violation (CPV) 
parameters''. The results  in 
eqs. (\ref{12R})-(\ref{23R})
imply that in the absence of significant 
RG effects, the ``double'' ratios
%%%%%%%%%%%%%%%%%%%%%%%%%%%%%%5
\begin{align}
\text{R}(21/31) \equiv \frac{\text{BR}(\mu \to e + \gamma)}
{\text{BR}(\tau \to e + \gamma)}~\text{BR}(\tau \to e\nu_{\tau}\bar{\nu}_e)\;,~ \text{R}(21/32) \equiv \frac{\text{BR}(\mu \to e + \gamma)}
{\text{BR}(\tau \to \mu + \gamma)}
~\text{BR}(\tau \to e\nu_{\tau}\bar{\nu}_e)\;,
\label{DoubleR}
\end{align}
%%%%%%%%%%%%%%%%%%%%%%%%%%%%%%%%%%%%%%%%%%%%%%%
%
depend in the region of validity of eqs. (\ref{eq_ijg}) 
and (\ref{eq_ms}) in the relevant SUSY 
parameter space, on the neutrino masses $m_j$,
mixing  angles $\theta_{12}$, 
$\theta_{23}$,  $\theta_{13}$ and
Majorana and Dirac CP-violation phases 
$\alpha$, $\beta_M$ and $\delta$
at $\sim M_Z$, 
as well as on the leptogenesis 
CP-violating (CPV) parameters 
$a$, $b$ and $c$. The dependence of 
the Dirac phase $\delta$
can be significant only if the CHOOZ angle
$\theta_{13}$ is sufficiently large.
The case of real matrix $\mathbf{R} \neq {\bf 1}$ 
corresponds to 
$a=0$, $b=0$ and $c=0$. 
Although the general expressions for
$(Y_{\nu}^{\dagger}Y_{\nu})_{ij}$, $i\neq j$,
eqs. (\ref{12R})-(\ref{23R}),
include several terms, there are few
physically interesting cases in which 
the expressions simplify and
the dependence on the Majorana
CP-violation phase(s) and/or 
on the leptogenesis CPV parameters is prominent. 

%%%%%%%%%%%%%%%%%%%%%%%%%%%%%%%
%
\subsection{\label{sec:NH}\large{Normal Hierarchical Neutrino Mass Spectrum}}
%
%%%%%%%%%%%%%%%%%%%%%%%%%%%%%%%
%

\hskip 1.0truecm  If the neutrino mass spectrum
is of the {\it normal hierarchical} (NH) type
and $m_1$ is negligibly small, i.e., 
$|a|\sqrt{m_1m_2},|b|\sqrt{m_1m_3} \ll |c|\sqrt{m_2m_3}$,
the quantities 
$(Y_{\nu}^{\dagger}Y_{\nu})_{ij}$, $i\neq j$, depend, 
%as eqs. (\ref{12R}) - (\ref{23R}) show,
in particular, on the same Majorana phase difference
$(\alpha - \beta_M)$ on which the effective Majorana mass,
eq. (\ref{meffNH2}), depends, on the Dirac CPV phase 
$\delta$ and one leptogenesis CPV parameter, $c$. 
The terms $\propto c\sqrt{m_2m_3}s_{13}e^{-i\delta}$ 
give always subdominant contributions 
in $|(Y_{\nu}^{\dagger}Y_{\nu})_{12,13}|$.
For $s_{13} \ll \tan\theta_{12} \sim 0.65$ 
they are negligible. In this case the expressions for 
$|(Y_{\nu}^{\dagger}Y_{\nu})_{12,13}|$ simplify:
%%%%%%%%%%%%%%%%%%%%%%%%%%%%%%%%%%%%
\begin{align}
\label{12RNH}
|(Y_{\nu}^{\dagger}Y_{\nu})^{\rm NH}_{12}|^2&\cong
\frac{M_R^2}{v_u^4}~
\left |c_{23}~P^{\rm NH} + s_{23}~Q^{\rm NH}\right |^2 \;,\\%~~{\rm NH}\;,\\
\label{13RNH}
|(Y_{\nu}^{\dagger}Y_{\nu})^{\rm NH}_{13}|^2& \cong
\frac{M_R^2}{v_u^4}~
\left |-s_{23}~P^{\rm NH} + c_{23}~Q^{\rm NH}\right |^2\;,% \\
\end{align}
%%%%%%%%%%%%%%%%%%%%%%%%%%%%%%%%%%%%%%%%%%%%%
%
where 
%%%%%%%%%%%%%%%%%%%%%%%%%%%%%%%%%%%%
\begin{align}
\label{NHP}
P^{\rm NH} & = (\deltasol)^{\frac{1}{2}}c_{12}s_{12}\;,\\%~~~
\label{NHQ}
Q^{\rm NH} & = (\deltaatm)^{\frac{1}{2}}s_{13}e^{-i\delta} + 
i~2c~(\deltasol \deltaatm)^{\frac{1}{4}}~
s_{12}~e^{i\frac{\alpha-\beta_M}{2}}\;. 
% \label{NHPQ}
\end{align}
%%%%%%%%%%%%%%%%%%%%%%%%%%%%%%%%%%%%
% 
The double ratio $\text{R}(21/31)$
is determined completely by the solar and atmospheric 
neutrino oscillation parameters 
$\deltasol$, $\theta_{12}$ and $\deltaatm$ and $\theta_{23}$,
by the CHOOZ angle $\theta_{13}$ and by the
Majorana and Dirac
CPV phases $(\alpha - \beta_M)$ and $\delta$ and 
by the leptogenesis CPV parameter $c$:
%%%%%%%%%%%%%%%%%%%%%%%%%%%%%%%%%%%%
\begin{align}
\text{R}(21/31)\cong 
\frac{\left |c_{23}~P^{\rm NH} + s_{23}~Q^{\rm NH}\right |^2}
{\left |-s_{23}~P^{\rm NH} + c_{23}~Q^{\rm NH}\right |^2}
\;.
\label{R2131NH0} 
\end{align}
%%%%%%%%%%%%%%%%%%%%%%%%%%%%%%%%%%%%%%%%%%%%%
%
where $P^{\rm NH}$ and $Q^{\rm NH}$
are given by eqs. (\ref{NHP}) and (\ref{NHQ}).
It follows from eqs. (\ref{NHP}), (\ref{NHQ}) and
(\ref{R2131NH0}) that if $(\alpha - \beta_M) = 0$ and
$\delta = \pm \pi/2,\pm 3\pi/2$,
we would have
%%%%%%%%%%%%%%%%%%%%%%%%%%%%%%%%%%%%
\begin{align}
\text{R}(21/31)\cong 
\frac{c_{23}^2 \left |P^{\rm NH}\right |^2 + 
s_{23}^2\left |Q^{\rm NH}\right |^2}
{s_{23}^2 \left |P^{\rm NH}\right |^2 + 
c_{23}^2\left |Q^{\rm NH}\right |^2}\;.
\label{R2131NH1} 
\end{align}
%%%%%%%%%%%%%%%%%%%%%%%%%%%%%%%%%%%%%%%%%%%%%
%
\noindent For the best fit value
$\sin^22\theta_{23} = 1$ we get $\text{R}(21/31) = 1$ 
independently  of the value of 
the leptogenesis CPV parameter
$c$, although the corresponding branching
ratios $\text{BR}(\mu \to e + \gamma)$ and
$\text{BR}(\tau \to e + \gamma)$ can exhibit strong 
dependence on $c$. If $\theta_{23}$ differs 
somewhat from $\pi/4$, the dependence of
$\text{R}(21/31)$ on $c$ will, in general,
be relatively mild. For  
$|P^{\rm NH}|^2 \gg |Q^{\rm NH}|^2$ 
($|P^{\rm NH}|^2 \ll |Q^{\rm NH}|^2$), however, 
the dependence of $\text{R}(21/31)$ on $c$ 
will be negligible even if $\theta_{23} \neq \pi/4$
and we would have 
$\text{R}(21/31) \cong \cot^2\theta_{23}~(\tan^2\theta_{23})$.

 For $|c| \ltap 0.1$ and $s_{13}$ having a value close to the 
existing (3$\sigma$) upper limit $s_{13} \cong 0.2$,  
the term $\propto c$ in $Q^{\rm NH}$, eq. (\ref{NHQ}), 
gives practically negligible
contributions in $|(Y_{\nu}^{\dagger}Y_{\nu})_{12,13}|$,
while $P^{\rm NH}$, eq. (\ref{NHP}), gives a subleading
but non-negligible
contribution. Correspondingly, the double ratio
$\text{R}(21/31)$ exhibits in this case a significant dependence
on the Dirac phase $\delta$ and is essentially independent 
of the leptogenesis CPV parameter $c$ and
the Majorana phase $(\alpha - \beta_M)$ (Fig. 1).  

   If $s_{13} \ltap 0.1$, but $s_{13}$ is 
not much smaller than 
$\sqrt{\deltasol}c_{12}s_{12}/\sqrt{\deltaatm}\sim 0.07$,
the branching ratios $\text{BR}(\mu \to e + \gamma)$
and $\text{BR}(\tau \to e + \gamma)$ 
still depend on the CHOOZ mixing angle $\theta_{13}$
and the phase $\delta$.  For $ 0.01 \ltap |c| \ltap 0.10$,
$\text{BR}(\mu \to e + \gamma)$
and $\text{BR}(\tau \to e + \gamma)$ can 
exhibit significant dependence also
on the Majorana phase $(\alpha - \beta_M)$. 
The dependence of the double ratio 
$\text{R}(21/31)$ on $(\alpha - \beta_M)$ 
and $\delta$ can be very strong due to possible
mutual compensation between $P^{\rm NH}$ and $Q^{\rm NH}$
(see eq. (\ref{R2131NH0})). For 
$(\alpha - \beta_M) \cong 0$, $s_{13} = 0.10$ and 
sufficiently small $|c|$, % $c=0$, 
for instance, we can have  $\text{R}(21/31) \sim 10^{-2}$ or
$\text{R}(21/31) \sim 10^{2}$  depending on 
whether $\delta \cong \pi$ or $\delta \cong 0$;
for $(\alpha - \beta_M) \cong \pi$ and
$|c|\sim 0.03$, $\text{R}(21/31)$ can have a value 
$\text{R}(21/31) \sim 10^{-3}$ or
$\text{R}(21/31) \sim 10^{3}$, respectively (Fig. 1).

    For rather small values of $s_{13}$, namely, 
$s_{13}\ll \sqrt{\deltasol}c_{12}s_{12}/\sqrt{\deltaatm}$, 
the dependence of $|(Y_{\nu}^{\dagger}Y_{\nu})_{12,13}|$ on
$s_{13}e^{-i\delta}$ is insignificant and can be neglected.
Under the latter condition we also have
$\sqrt{\deltaatm}s^2_{13}\ll \sqrt{\deltasol}s^2_{12}$.
The effective  Majorana mass in $\betabeta$-decay is given
correspondingly by $\meff_{\rm NH} \cong \sqrt{\deltasol}s_{12}^2$.
The quantities $P^{\rm NH}$ and $Q^{\rm NH}$
can be written as:
%%%%%%%%%%%%%%%%%%%%%%%%%%%%%%%%%%%%
\begin{align}
% \label{NHP}
P^{\rm NH}\cong (\meff_{\rm NH})^{\frac{1}{2}}
(\deltasol)^{\frac{1}{4}}c_{12}\;,
% \label{NHQ}
~Q^{\rm NH}\cong i~2c~(\meff_{\rm NH})^{\frac{1}{2}}
(\deltaatm)^{\frac{1}{4}}~
e^{i\frac{\alpha-\beta_M}{2}}\;. 
\label{NHPQ1}
\end{align}
%%%%%%%%%%%%%%%%%%%%%%%%%%%%%%%%%%%%
%
Thus, in this case $\text{BR}(\mu \to e + \gamma)\propto \meff_{\rm NH}$
and $\text{BR}(\tau \to e + \gamma)\propto \meff_{\rm NH}$. 
Given $\deltasol$, $\deltaatm$,
$\theta_{12}$ and $\theta_{23}$,
the ratio $\text{R}(21/31)$ is 
determined by the Majorana phase difference 
$(\alpha - \beta_M)$ and the leptogenesis CPV 
parameter $c$: 
%%%%%%%%%%%%%%%%%%%%%%%%%%%%%%%%%%%%
\begin{align}
\label{R2131NH1}
\text{R}(21/31) \cong \frac{
\left |(\deltasol)^{\frac{1}{4}}c_{12}\cot\theta_{23} + 
i~2c~(\deltaatm)^{\frac{1}{4}}~e^{i\frac{\alpha-\beta_M}{2}}
\right |^2}
{\left |(\deltasol)^{\frac{1}{4}}c_{12} - 
i~2c~(\deltaatm)^{\frac{1}{4}}~
e^{i\frac{\alpha-\beta_M}{2}}\cot\theta_{23}~
\right |^2} \;.
\end{align}
%%%%%%%%%%%%%%%%%%%%%%%%%%%%%%%%%%%%%%%%%%%%%
%
Obviously, for $(\alpha-\beta_M) \cong 0$, we have
$\text{R}(21/31) \cong 1$. If, however, 
$(\alpha-\beta_M) \cong \pm \pi$,
the double ratio $\text{R}(21/31)$
can depend strongly on the value of $|c|$, provided
$|c| \gtap 0.05$. For 
$2|c| \sim (\deltasol/\deltaatm)^{\frac{1}{4}} \cong 0.42$,
the two terms in the numerator (denominator) 
% in eq. (\ref{R2131NH1})
of the expression for $\text{R}(21/31)$ 
can compensate (partially) each other and one can have 
$\text{R}(21/31) \sim (10^{-3} - 10^{-2})$ or
$\text{R}(21/31) \sim (10^{3} - 10^{2})$
depending on the sign of $c$ (Fig. 1).
 
\begin{figure}
\begin{tabular}{cc}
\includegraphics[scale=1.25]{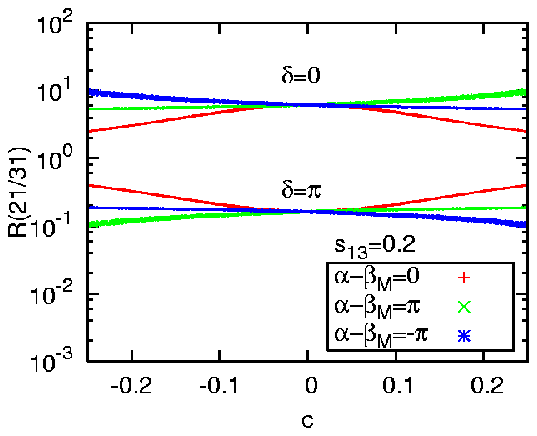}&
\includegraphics[scale=1.25]{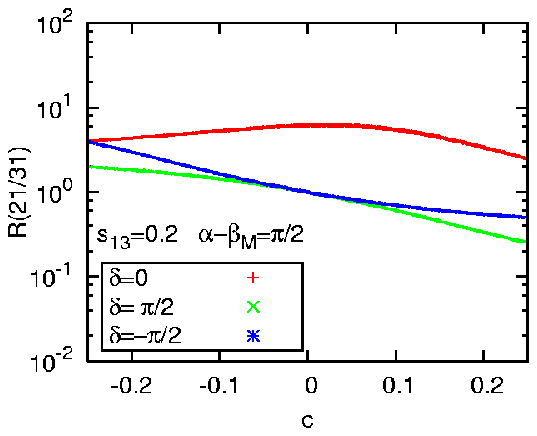}\\
(a)&(b)\\
\includegraphics[scale=1.25]{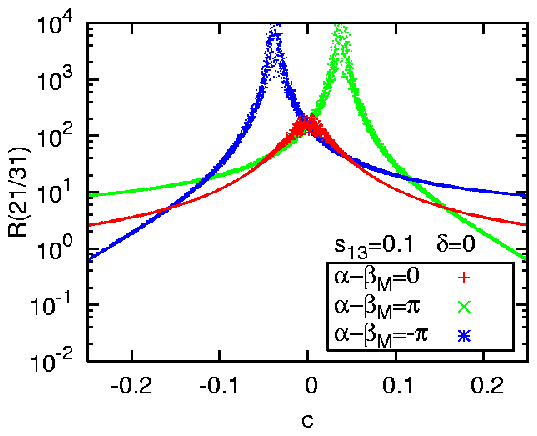}&
\includegraphics[scale=1.25]{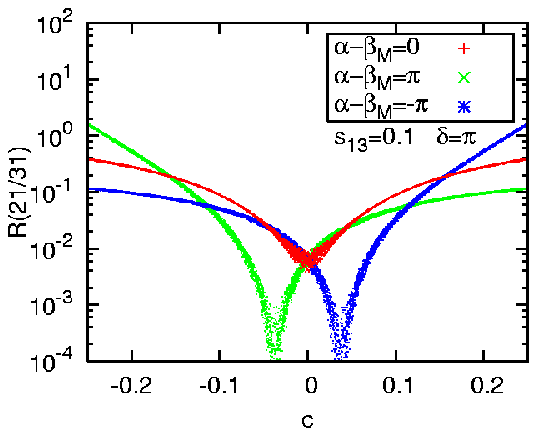}\\
(c)&(d)\\
\includegraphics[scale=1.25]{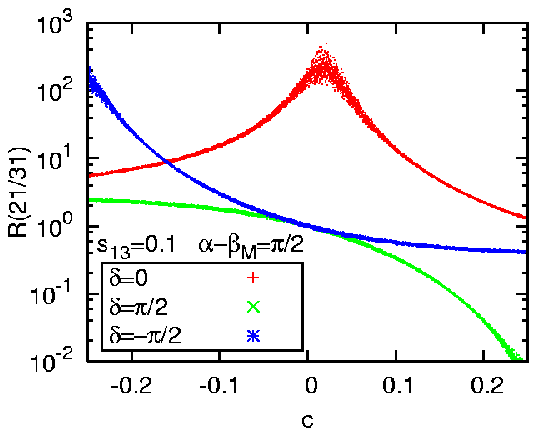}&
\includegraphics[scale=1.25]{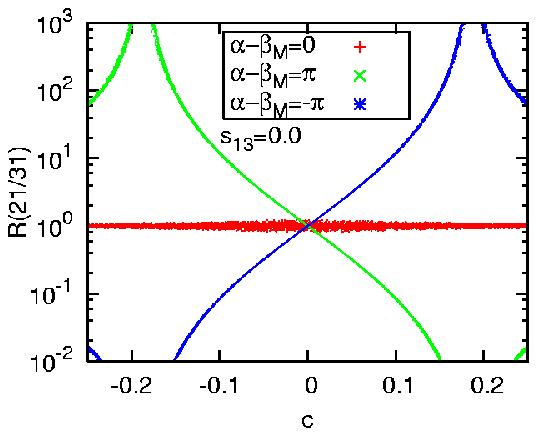}\\
(e)&(f)\\
\end{tabular}
\caption{The double ratio $\text{R}(21/31)$ %, eq. (\ref{DoubleR}), 
in the case of NH light neutrino mass spectrum,
as a function of the leptogenesis CPV parameter
$c$, for $s_{13} = 0.2;~0.1;~0$ and
several characteristic values of the Dirac and 
Majorana CPV phases $\delta$ and $\alpha-\beta_M$. 
The figure was obtained using the best fit values 
of the solar and atmospheric 
neutrino oscillation parameters 
$\deltasol$, $\sin^2\theta_{12}$,
$\deltaatm$ and $\sin^22\theta_{23}$.
The lightest neutrino mass $m_1$ was set to 0.
The effects of the higher order corrections 
in leptogenesis CPV parameters is reflected 
in the width of the lines
(see text for further details). 
}
\label{NH-R2131-alp}
\end{figure}

 If $|c|$ is relatively small, 
$2|c| \ll (\deltasol/\deltaatm)^{\frac{1}{4}}\cong 0.42$,
$\text{BR}(\mu \to e + \gamma)$ and 
$\text{BR}(\tau \to e + \gamma)$  
are practically independent of $c$ and $(\alpha - \beta_M)$.
This case was analised recently in \cite{PShinYasu05}.
If, for instance, 
% $s_{13}$ is rather small, 
$s_{13}\ll \sqrt{\deltasol}c_{12}s_{12}/\sqrt{\deltaatm}$,
we find from  eq. (\ref{R2131NH1}): 
%%%%%%%%%%%%%%%%%%%%%%%%%%%%%%
\begin{align}
\text{R}(21/31) \cong \cot^{2}\theta_{23}\;.
% \tan^{-2}\theta_{23}\;.
\label{R2131NH}
\end{align}
%%%%%%%%%%%%%%%%%%%%%%%%%%%%%%
%
\hskip 1.0truecm The results for the double 
ratio $\text{R}(21/31)$ discussed above are
illustrated in Fig.~\ref{NH-R2131-alp}, where the dependence of  
$\text{R}(21/31)$ 
on the leptogenesis CPV parameter $c$
for $s_{13} = 0;~0.10;~0.20$ and few characteristic 
values of the Majorana and Dirac CPV phases 
$(\alpha-\beta_M) = 0;\pi/2;\pm \pi$ and
$\delta =0;\pm \pi/2;\pm \pi$ are shown.
The figure was obtained using the best fit values 
of the solar and atmospheric 
neutrino oscillation parameters $\theta_{12}$,
$\deltasol$, $\theta_{23}$ and $\deltaatm$. 
The lightest neutrino mass $m_1$ was set to 0.
The quantities $|(Y_{\nu}^{\dagger}Y_{\nu})^{\rm NH}_{12,13}|^2$
were calculated using eqs. (\ref{YnudYnu}) and 
(\ref{RPPY}) and not the approximate
expressions given in eqs. (\ref{12R}) and
(\ref{13R}). The leptogenesis CPV violating 
parameters $a$ and $b$ can contribute only 
to the higher order corrections 
$\mathcal{O}(r^2,s_{13}^2)$ in 
$|(Y_{\nu}^{\dagger}Y_{\nu})^{\rm NH}_{12,13}|^2$. 
These corrections can be relevant for the evaluation of
$\text{R}(21/31)$ in the case of cancellations between
the terms in $|(Y_{\nu}^{\dagger}Y_{\nu})^{\rm NH}_{12,13}|^2$,
which provide the leading order contributions.
We have allowed $a$ and $b$ to vary in the same 
interval as the parameter $c$ in the calculations.
The effects of the higher order corrections 
due to $a$ and $b$ is reflected
in the widths of the lines in Fig. 1.

  We shall perform next similar analysis for the
double ratio $\text{R}(21/32) = 
\text{BR}(\mu \to e + \gamma)/\text{BR}(\tau \to \mu + \gamma)$.
As can be easily verified using eqs. (\ref{12R}) and (\ref{23R}) and 
the known values of the neutrino oscillation parameters, 
for $|c|\leq 0.3$ we always have 
%%%%%%%%%%%%%%%%%%%%%%%%%%%%%%%%%%%%
\begin{align}
\text{R}(21/32) <  1\;.
\label{R2132NH00} 
\end{align}
%%%%%%%%%%%%%%%%%%%%%%%%%%%%%%%%%%%%%%%%%%%%%
%
\noindent  Typically 
% (in most of the allowed 
% region of the space of parameters) 
the stronger inequality $\text{R}(21/32) \ll 1$ holds 
\footnote{This is in contrast to the case of 
normal hierarchical heavy Majorana neutrino
mass spectrum, in which one typically has 
$\text{R}(21/32) \sim  1$ \cite{PShinYasu05}.}
(see further).

 It is not difficult to convince oneself also that 
the term $\propto \Delta_{31}s_{23}c_{23}$
dominates in $|(Y_{\nu}^{\dagger}Y_{\nu})_{23}|^2$.
Indeed, we have $(\deltasol/\deltaatm)^{\frac{1}{2}} \cong 0.18$,
$c_{12}|\cos\theta_{23}| \ltap 0.24$,
$s_{13}s_{12}\sin2\theta_{23} \ltap 0.12$, and 
for $|c|\leq 0.2~(0.3)$, the terms $\propto c$ 
in eq. (\ref{23R}) give contributions which do not exceed
approximately 8\% (18\%). Keeping only the largest
of these contributions we have:
$|(Y_{\nu}^{\dagger}Y_{\nu})_{23}|^2
\cong M^2_R v_u^{-4}~|\sqrt{\deltaatm}s_{23}c_{23}
+ 2ic~(\deltasol \deltaatm)^{\frac{1}{4}}c_{12}
\cos(\alpha -\beta_M)/2|^2$. Thus, 
the branching ratio $\text{BR}(\tau \to \mu + \gamma)$
exhibits very weak dependence on $c$ and $(\alpha -\beta_M)$.
Up to the indicated corrections which for $|c| \leq 0.3$
can increase  $|(Y_{\nu}^{\dagger}Y_{\nu})_{23}|^2$ 
by not more than 18\%, we have:
%%%%%%%%%%%%%%%%%%%%%%%%%%%%%%%
\begin{align}
\label{23RNH}
|(Y_{\nu}^{\dagger}Y_{\nu})_{23}|^2 
\cong \frac{M^2_R}{v^4_u}~\deltaatm s^2_{23} c^2_{23}
\;.
\end{align}
%%%%%%%%%%%%%%%%%%%%%%%%%%%%%%%%%%%%%%%%%%%%%
%
Thus, for $|c| \ltap 0.3$ in the case under discussion,
$\text{BR}(\tau \to \mu + \gamma)$ 
depends essentially only on the atmospheric 
neutrino oscillation parameters 
$\deltaatm$ and  $\theta_{23}$
(and not on the Dirac and Majorana CPV phases,
leptogenesis CPV parameters or solar 
neutrino oscillation parameters 
$\deltasol$ and $\theta_{12}$)
and has a relatively simple form:
$\text{BR}(\tau \to \mu + \gamma) \cong 
\text{F} \times (\deltaatm/(4v^2_u))~\sin^22\theta_{23}$,
where the factor $\text{F} \propto M_R^2/v^2_u$ contains all the
dependence on $M_R$, $\tan\beta$ and 
the SUSY breaking parameters (see eq. (\ref{eq_ijg})).
The double ratio  $\text{R}(21/32)$, 
however, depends in the case under discussion both on 
$c~e^{i\frac{\alpha-\beta_M}{2}}$ and $s_{13}e^{-i\delta}$: 
%%%%%%%%%%%%%%%%%%%%%%%%%%%%%%%%%%%%
\begin{align}
\text{R}(21/32)\cong 
\frac{\left |c_{23}~P^{\rm NH} + s_{23}~Q^{\rm NH}\right |^2}
{\deltaatm s^2_{23}c^2_{23}}
\;,
\label{R2132NH0} 
\end{align}
%%%%%%%%%%%%%%%%%%%%%%%%%%%%%%%%%%%%%%%%%%%%%
%
where $P^{\rm NH}$ and $Q^{\rm NH}$ are given by 
eqs. (\ref{NHP}) and (\ref{NHQ}).

  For $s_{13} \cong 0.2$ and $|c| \ltap 0.25$, 
we have $\sqrt{\deltaatm}s_{13} \cong 2.3\sqrt{\deltasol}c_{12}s_{12}$
and $\sqrt{\deltaatm}s_{13}
\gtap 1.7~(2c (\deltaatm \deltasol)^{\frac{1}{4}}s_{12})$.
The double ratio $\text{R}(21/32)$ exhibits noticeable dependence
on the CPV phases $(\alpha - \beta_{M})$ and $\delta$.
For  $(\alpha - \beta_{M})=0$ and $\delta=0$, the term 
$\propto c$ in $Q^{\rm NH}$ gives a subdominant contribution
and $\text{R}(21/32)$ is practically independent of $c$.
If $\delta = \pi$, however, the term  
$\propto \sqrt{\deltaatm}s_{13}$ in $Q^{\rm NH}$
can be compensated partially by $P^{\rm NH}$ and for 
sufficiently large values of $|c|$ the term
$\propto c$ in $Q^{\rm NH}$ can be non-negligible.
For $|c| \ltap 0.1$ in this case we can have 
$\text{R}(21/32)\sim {\rm few}\times 10^{-2}$, while if
$(\alpha - \beta_{M})= \pi$ and $|c| \cong 0.2$,
$\text{R}(21/32)$ can be as small as 
$\text{R}(21/32)\sim {\rm few}\times 10^{-3}$. 

   In the case of 
$s_{13}\sim \sqrt{\deltasol}c_{12}s_{12}/\sqrt{\deltaatm}\sim 0.07$,
partial compensation between the three 
terms in the numerator of the double ratio $\text{R}(21/32)$
can take place.  The double ratio 
%%%%%%%%%%%%%%%%%%%%%%%%%%%%%%%%%%%%%%%%%%%%
% \newpage \noindent 
$\text{R}(21/32)$ 
can be particularly strongly suppressed for 
$\delta \cong \pi$, when values of $\text{R}(21/32) 
\sim (10^{-3} - 10^{-4})$ for $|c| \sim 0.05$ are possible.
Similar mutual compensations between the terms in the 
numerator of $\text{R}(21/32)$ can be realised if 
$s_{13}\ll \sqrt{\deltasol}c_{12}s_{12}/\sqrt{\deltaatm}$
and $|c| \sim (0.15 - 0.20)$. One can have
$\text{R}(21/32) \sim (10^{-3} - 10^{-4})$ in this case as well.
For sufficiently small $s_{13}$ the dependence on the phase 
$\delta$ is obviously insignificant and we have:
%%%%%%%%%%%%%%%%%%%%%%%%%%%%%%%%%%%%
\begin{align}
\text{R}(21/32)\cong 
\frac{\meff_{\rm NH}}{\sqrt{\deltaatm}s_{23}^2c^2_{23}}
\left | \left (\frac{\deltasol}{\deltaatm} \right )^{\frac{1}{4}}c_{23}c_{12} 
+ i~2c~s_{23}e^{i\frac{\alpha-\beta_M}{2}}\right |^2 
\;.
\label{R2132NHs130} 
\end{align}
%%%%%%%%%%%%%%%%%%%%%%%%%%%%%%%%%%%%%%%%%%%%%
%
If in addition 
$2|c| \ll (\deltasol/\deltaatm)^{\frac{1}{4}}\cong 0.42$, 
$\text{R}(21/32)$ is also practically independent of
$c$ and  $(\alpha - \beta_M)$. It is determined completely 
by the solar and atmospheric neutrino oscillation
parameters:
%%%%%%%%%%%%%%%%%%%%%%%%%%%%%%%%%%%%
\begin{align}
\text{R}(21/32)\cong 
\meff_{\rm NH}~
\frac{(\deltasol)^{\frac{1}{2}}}{\deltaatm s_{23}^{2}}~c^2_{12} 
\cong \frac{\deltasol}{\deltaatm s_{23}^2}~c^2_{12}s^2_{12} 
\simeq 1.3\times 10^{-2} 
\;.
\label{R2132NH}
\end{align}
%%%%%%%%%%%%%%%%%%%%%%%%%%%%%%%%%%%%%%%%%%%%%
% 

\hskip 1.0truecm 
The specific features of the double ratio  $\text{R}(21/32)$
discussed above are evident in Fig. 2, where the dependence of 
$\text{R}(21/32)$ on the leptogenesis CPV parameter $c$, 
$|c| \leq 0.25$, for three values of $s_{13} = 0;~0.1;~0.2$, 
and several characteristic values of the 
Majorana and Dirac  CPV phases 
$(\alpha - \beta_M)$ and $\delta$ is shown.
Figure 2 was obtained using the same method 
and the same best fit values of the oscillation parameters
$\deltasol$, $\sin^2\theta_{12}$,
$\deltaatm$ and $\sin^22\theta_{23}$, as Fig. 1.
% (see the discussion above).
 %%%%%%%%%%%%%%%%%%%%%%%%%%%%%%%
%
\subsection{\label{sec:NH}\large{Inverted Hierarchical Spectrum}}
%
%%%%%%%%%%%%%%%%%%%%%%%%%%%%%%%
%

\hskip 1.0truecm  If neutrino mass spectrum is 
{\it inverted hierarchical} (IH) one has $m_3 \ll m_{1,2}$,
and we shall assume that the terms $\propto \sqrt{m_3}$ 
in eqs. (\ref{12R})-(\ref{23R}) can be neglected. 
For $\mefff_{\rm IH}$ we have
$\mefff_{\rm IH} \cong \sqrt{|\deltaatm|}(c_{12}^2 + s_{12}^2e^{i\alpha})$
(see eq. (\ref{meffIH1})).
Now  $|(Y_{\nu}^{\dagger}Y_{\nu})_{ij}|$, $i\neq j$,
depend on the Majorana phase $\alpha$, on the
leptogenesis CPV parameter $a$ and, if $s_{13}$
has a value close to the current upper limit 
- on the Dirac phase $\delta$:
%%%%%%%%%%%%%%%%%%%%%%%%%%%%%%%%%%%%
\begin{align}
\label{12RIH}
|(Y_{\nu}^{\dagger}Y_{\nu})^{\rm IH}_{12}| & \cong
~\frac{M_R}{v_u^2}~
\left |c_{23}~P^{\rm IH} + s_{23}~Q^{\rm IH}\right | \;,\\%~~{\rm NH}\;,\\
\label{13RIH}
|(Y_{\nu}^{\dagger}Y_{\nu})^{\rm IH}_{13}| & \cong
\frac{M_R}{v_u^2}~
\left |-s_{23}~P^{\rm IH} + c_{23}~Q^{\rm IH}\right | \;,\\
|(Y_{\nu}^{\dagger}Y_{\nu})^{\rm IH}_{23}| & \cong
\frac{M_R}{v_u^2}\sqrt{|\deltaatm|}~c_{23}s_{23}
\left |-1 + 4ac_{12}s_{12}\sin\frac{\alpha}{2}\right | \;,
\label{23RIHa}
\end{align}
%%%%%%%%%%%%%%%%%%%%%%%%%%%%%%%%%%%%%%%%%%%%%
%
where 
%%%%%%%%%%%%%%%%%%%%%%%%%%%%%%%%%%%%
\begin{align}
\label{IHP}
P^{\rm IH} = \frac{1}{2}~\frac{\deltasol}{\sqrt{|\deltaatm|}}~c_{12}s_{12} 
+ i~2a~\mefff_{\rm IH}~e^{-i\frac{\alpha}{2}}\;, \\% ~~~
Q^{\rm IH} = % (|\deltaatm)^{\frac{1}{2}}~s_{13}~e^{-i\delta}\;.
- \sqrt{|\deltaatm|}~s_{13}~e^{-i\delta}
\left (1 + 4ac_{12}s_{12}~\sin\frac{\alpha}{2}\right )\;. 
\label{IHPQ}
\end{align}
%%%%%%%%%%%%%%%%%%%%%%%%%%%%%%%%%%%%
% 
For $s_{13}$ satisfying
% is relatively small, 
%%%%%%%%%%%%%%%%%%%%%%%%%%%%%%%%%%%%%%%%%
\begin{align}
\sin\theta_{13}(1 + 2|a|\sin 2\theta_{12}) \ll 
{\rm min}\left (2|a|\cos2\theta_{12},   
\frac{\deltasol}{4|\deltaatm|}\sin 2\theta_{12} \right )
\label{s13IH}
\end{align}
%%%%%%%%%%%%%%%%%%%%%%%%%%%%
%
the dependence of $|(Y_{\nu}^{\dagger}Y_{\nu})^{\rm IH}_{12,13}|$
on the Dirac phase $\delta$ would be insignificant.
The terms $\propto Q^{\rm IH}$ in
eqs. (\ref{12RIH}) and (\ref{13RIH}) are negligible,
and  the ratio of $\text{BR}(\mu \to e + \gamma)$
and $\text{BR}(\tau \to e + \gamma)$ is given by
%%%%%%%%%%%%%%%%%%%%%%%%%%%%%%
\begin{align}
\text{R}(21/31) \cong \cot^2\theta_{23}\;,
\label{R2131IHa}
\end{align}
%%%%%%%%%%%%%%%%%%%%%%%%%%%%%%%%%%%%%%%%%%%%%%%
%
independently of the values of the
Majorana CPV phase $\alpha$, leptogenesis 
CPV parameter $a$, etc. (Fig. 3). If in addition 
$|a| \ll (\deltasol/(8|\deltaatm|))\sin 2\theta_{12}\cong 3.6\times 10^{-3}$,
$\text{BR}(\mu \to e + \gamma)$ and $\text{BR}(\tau \to e + \gamma)$
also will not depend on $\alpha$ and $a$:\\
$|(Y_{\nu}^{\dagger}Y_{\nu})^{\rm IH}_{12(13)}|^2 \cong C^2_{12(13)}~
(M^2_R \deltasol/v_u^4)(\deltasol/(16|\deltaatm|))\sin^22\theta_{12}$,
where $C_{12(13)} \equiv c_{23}~(s_{23})$.
In the case of $|a|\cos2\theta_{12} \gg (\deltasol/(8|\deltaatm|))
\sin 2\theta_{12}\cong 4\times 10^{-3}$, however, we have:
%%%%%%%%%%%%%%%%%%%%%%%%%%%%%%%%%%%%
\begin{align}
|(Y_{\nu}^{\dagger}Y_{\nu})^{\rm IH}_{12(13)}| \cong
2|a|~\left | \mefff_{\rm IH} \right |
\frac{M_R}{v_u^{2}}~C_{12(13)}\;.
\label{Y1213IH}
\end{align}
%%%%%%%%%%%%%%%%%%%%%%%%%%%%%%%%%%%%%%%%%%%%%
%
Thus, both $\text{BR}(\mu \to e + \gamma)$ 
and $\text{BR}(\tau \to e + \gamma)$ are proportional to 
%$\text{BR}(\mu(\tau) \to e + \gamma) \propto$ 
$|a|^2~|\mefff_{\rm IH}|^2$. 

   We get $\text{R}(21/31) \sim 1$ also when 
$s_{13} \gg (\deltasol/(4|\deltaatm|))
\sin 2\theta_{12}\cong 8\times 10^{-3}$, provided
$\alpha \cong 0$ and $\delta \cong 0;~\pi$ (Fig. 3).
In this case $|\mefff_{\rm IH}|^2 \cong |\deltaatm|$,
$|(Y_{\nu}^{\dagger}Y_{\nu})^{\rm IH}_{12}| \cong
(4a^2c^2_{23} + s^2_{13}s^2_{23})|\deltaatm|M^2_R/v_u^4$,
$|(Y_{\nu}^{\dagger}Y_{\nu})^{\rm IH}_{13}| \cong
(4a^2s^2_{23} + s^2_{13}c^2_{23})|\deltaatm|M^2_R/v_u^4$ and
%%%%%%%%%%%%%%%%%%%%%%%%%%%%%%%%%%%%
\begin{align}
\text{R}(21/31)\cong 
\frac{4a^2c^2_{23} + s^2_{13}s^2_{23}}
{4a^2s^2_{23} + s^2_{13}c^2_{23}}\;.
\label{R2131IHb}
\end{align}
%%%%%%%%%%%%%%%%%%%%%%%%%%%%%%%%%%%%%%%%%%%%%
% 
If, however, $\alpha$ is significantly 
different from zero, say $\alpha \cong \pm \pi/2;~\pm \pi$,
and $|a|$ is sufficiently large, being comparable 
in magnitude to $s_{13}$,
the terms $\propto P^{\rm IH}$ and $\propto Q^{\rm IH}$ 
in $|(Y_{\nu}^{\dagger}Y_{\nu})^{\rm IH}_{12}|$
or $|(Y_{\nu}^{\dagger}Y_{\nu})^{\rm IH}_{13}|$
can partially compensate each other and we can have 
$\text{R}(21/31)\sim (10^{-3} - 10^{-2})$ or
$\text{R}(21/31)\sim (10^{2} - 10^{3})$ (Fig. 3).
For given $|a|$ and $s_{13}$, the degree of 
compensation depends on the values of $\alpha$ 
and $\delta$ and on the ${\rm sgn}(a)$.
It is maximal in $|(Y_{\nu}^{\dagger}Y_{\nu})^{\rm IH}_{12(13)}|$,
for, e.g., $\alpha = \pi$ and $\delta = 0$, 
or $\alpha = -\pi$ and $\delta = \pi$,
and $a >0$ ($a <0$) (Fig. 3).

 The ratio of $\text{BR}(\mu \to e + \gamma)$
and $\text{BR}(\tau \to \mu + \gamma)$
depends both on $a$ and $\alpha$. For $|a| \leq 0.3$
we have $\text{R}(21/32)\ltap 1$; 
if $|a| \leq 0.1$, the stronger inequality 
$\text{R}(21/32)\ll 1$ typically
holds. For $|a| \ll (\deltasol/(8|\deltaatm|))\sin2\theta_{12} 
\cong 4\times 10^{-3}$ and negligibly small $s_{13}$, 
for instance, one finds \cite{PShinYasu05} 
$\text{R}(21/32) \cong 10^{-4}$.
If, however, the term $\propto a$ dominates in  
$|P^{\rm IH}|$, i.e., if $|a|\cos 2\theta_{12} 
\gg 4\times 10^{-3}$, we get (for $s_{13}\sim 0$) 
$\text{BR}(\mu \to e + \gamma) \propto |a|^2~|\mefff_{\rm IH}|^2$
and correspondingly,
%%%%%%%%%%%%%%%%%%%%%%%%%%%%%%%%%%%%
\begin{align}
\text{R}(21/32)\cong 
4~|a|^2 s^{-2}_{23}~r_{\rm IH}
\left | -1 + 2a\eta \left (1 - r_{\rm IH}
 \right )^{\frac{1}{2}} \right|^{-2}\;,
\label{R2132IHa}
\end{align}
%%%%%%%%%%%%%%%%%%%%%%%%%%%%%%%%%%%%%%%%%%%%%
%
where $\eta \equiv {\rm sgn}(\sin2\theta_{12}\sin\frac{\alpha}{2})$ and 
%%%%%%%%%%%%%%%%%%%%%%%%%%%%%%%%%%%%
\begin{align}
r_{\rm IH} \equiv \frac{(\meff_{\rm IH})^2}{|\deltaatm|} =
1 - \sin^22\theta_{12}\sin^2\frac{\alpha}{2}\;.
\label{R2132IHar}
\end{align}
%%%%%%%%%%%%%%%%%%%%%%%%%%%%%%%%%%%%%%%%%%%%%
% 
Now $\text{R}(21/32)$ can be considerably larger:
for $\alpha$ varying between 0 and $\pi$
and $|a|$ having a value, e.g., in the interval
(0.04 - 0.10), the ratio of interest
satisfies $1.9\times 10^{-3} \ltap 
\text{R}(21/32) \ltap 8.0\times 10^{-2}$,
the maximal value corresponding to
$|a|= 0.1$ and $\alpha = 0$.
%%%%%%%%%%%%%%%%%%%%%%%%%%%%%%%%%%%%%%%%%%% 

 The predictions for the double ratios $\text{R}(21/31)$ 
and $\text{R}(21/32)$, corresponding to 
IH light neutrino mass spectrum 
are illustrated in Figs. 3 and 4, respectively.
As in the case of Figs. 1 and 2,
the quantities $|(Y_{\nu}^{\dagger}Y_{\nu})^{\rm IH}_{ij}|^2$, $i\neq j$,
have been calculated using eqs. (\ref{YnudYnu}) and 
(\ref{RPPY}) rather than the approximate
expressions given in eqs. (\ref{12R}) and
(\ref{13R}). The lightest neutrino mass $m_3$ set to 0.
The leptogenesis CPV parameters
$b$ and $c$, which can contribute only
to the higher order corrections in 
$|(Y_{\nu}^{\dagger}Y_{\nu})^{\rm IH}_{ij}|^2$ of interest,
were varied in the same interval as the parameter 
$a$ in the calculations.
The effects of the higher order corrections 
due to $b$ and $c$ is reflected
in the widths of the lines in Figs. 3 and 4.

%%%%%%%%%%%%%%%%%%%%%%%%%%%%%%%
%
\subsection{\label{sec:NH}\large{Quasi-Degenerate Neutrinos}}
%
%%%%%%%%%%%%%%%%%%%%%%%%%%%%%%%
%
\hskip 1.0truecm For QD light neutrino mass spectrum, 
$m_{1,2,3} \cong m\gtap 0.1$ eV,
one has $\mefff_{\rm QD} \cong m(c_{12}^2 + s_{12}^2e^{i\alpha})$,
and $\sqrt{m_im_j} \cong m$ in eqs. (\ref{12R}) - (\ref{23R}).  
Barring ``accidental'' cancellations, we always have 
$|(Y_{\nu}^{\dagger}Y_{\nu})_{12}| \sim 
|(Y_{\nu}^{\dagger}Y_{\nu})_{13}|$,
and correspondingly $\text{BR}(\mu \to e + \gamma)\sim
\text{BR}(\tau \to e + \gamma)$, in this case (Fig. 5).
The expressions for $|(Y_{\nu}^{\dagger}Y_{\nu})_{12(13)}|$
of interest simplify if ${\rm max}(|a|,|b|,|c|)\gg 
{\rm max}(\deltasol/(4m^2),|\deltaatm|s_{13}/(4m^2))$ and 
$s_{13} \ltap 0.1$:
%%%%%%%%%%%%%%%%%%%%%%%%%%%%%%%%%%%%
\begin{align}
\label{12RQD}
|(Y_{\nu}^{\dagger}Y_{\nu})^{\rm QD}_{12}|&\cong
2~\frac{M_R}{v_u^2}~
\left |c_{23}~P^{\rm QD} + s_{23}~Q^{\rm QD}\right | \;,\\%~~{\rm QD}\;,\\
\label{13RQD}
|(Y_{\nu}^{\dagger}Y_{\nu})^{\rm QD}_{13}|& \cong
2~\frac{M_R}{v_u^2}~
\left |-s_{23}~P^{\rm QD} + c_{23}~Q^{\rm QD}\right |\;,
\end{align}
%%%%%%%%%%%%%%%%%%%%%%%%%%%%%%%%%%%%%%%%%%%%%
%
where 
%%%%%%%%%%%%%%%%%%%%%%%%%%%%%%%%%%%%
\begin{align}
\label{QDP}
P^{\rm QD} &= a~\mefff_{\rm QD}~e^{-i\frac{\alpha}{2}}\;,\\
\label{QDQ}
Q^{\rm QD} &= m \left [ \left (bc_{12} + cs_{12}e^{i\frac{\alpha}{2}}\right )
e^{-i\frac{\beta_M}{2}} 
+ ias_{13}\sin2\theta_{12}~e^{-i\delta}\sin \frac{\alpha}{2} \right ] 
\;. 
\end{align}
%%%%%%%%%%%%%%%%%%%%%%%%%%%%%%%%%%%%
% 
The condition specified above is compatible with the
leptogenesis constraints on the product $|abc|$ \cite{PPY03}.
For $\alpha \cong 0$ and $\beta_M \cong \pm \pi$ we get:
%%%%%%%%%%%%%%%%%%%%%%%%%%%%%%%%%%%%
\begin{align}
\text{R}(21/31)\cong 
\frac{c_{23}^2 \left |P^{\rm QD}\right |^2 + 
s_{23}^2\left |Q^{\rm QD}\right |^2}
{s_{23}^2 \left |P^{\rm QD}\right |^2 + 
c_{23}^2\left |Q^{\rm QD}\right |^2}\;.
\label{R2131QD1} 
\end{align}
%%%%%%%%%%%%%%%%%%%%%%%%%%%%%%%%%%%%%%%%%%%%%
%
Obviously, in this case either $\text{R}(21/31)\cong 1$ 
independently of the value of $\theta_{23}$, or 
$\text{R}(21/31)\cong \tan^2\theta_{23}~{\rm or}~\cot^2\theta_{23}$.

  Under the condition leading to eqs. 
(\ref{12RQD}) - (\ref{QDQ}), the quantity
$|(Y_{\nu}^{\dagger}Y_{\nu})_{23}|$, eq. (\ref{23R}),
cannot be simplified. The term  
$\propto \Delta_{31}$ in eq. (\ref{23R}) 
will be the dominant one if \\ 
${\rm max}(|a\sin(\alpha/2)|,2|a|s_{13},|a|^2),|b|,|c| \ll 
|\deltaatm|/(4m^2)$.
Given the leptogenesis constraint on $|abc|$, 
this is realised, e.g., for $m = 0.1$ eV if
$|a|\sim |b|\sim |c|\cong 10^{-2}$, 
or if $\sin(\alpha/2)\cong 0$, $s_{13} \cong 0$ and
$|a| \gg |b|,|c|$ but $|a|^2 \ll |\deltaatm|/(4m^2)$. 
In both cases we have
%%%%%%%%%%%%%%%%%%%%%%%%%%%%%%%%%%%%
\begin{align}
\text{R}(21/32)\cong 
\frac{16~m^4~|a|^2}{(\deltaatm)^2}
\left | \frac{\mefff_{\rm QD}}{m~s_{23}}
+ \frac{bc_{12} + cs_{12}e^{i\frac{\alpha}{2}}}{a~c_{23}}
e^{-i\frac{\beta_M - \alpha}{2}}\; \right |^2 \ll 1\;.
\label{R2132QD0}
\end{align}
%%%%%%%%%%%%%%%%%%%%%%%%%%%%%%%%%%%%%%%%%%%%%
%
For $m = 0.10$ eV and $|a| = |b| = |c| = 10^{-2}$,
the ratio $\text{R}(21/32)$ given by eq. (\ref{R2132QD0})
depends on $\alpha$, $\beta_M$,
${\rm sgn}(b/a)$ and ${\rm sgn}(c/a)$ and
satisfies $2\times 10^{-4} \ltap \text{R}(21/32) 
\ltap 3\times 10^{-1}$. If, however, 
$\alpha \cong 0$, $s_{13} \cong 0$ and 
$|a|\cong 0.2$ with  $|abc| \cong 10^{-5}$, the ``corrections''
$\propto |a|^2$ in eq. (\ref{23R}) will be non-negligible
since $|a|^2 \sim |\deltaatm|/(4m^2)$.
In this case we can have even $\text{R}(21/32) \cong 200$ 
as a consequence of rather strong partial 
cancellation between the different terms in 
the expression for $|(Y_{\nu}^{\dagger}Y_{\nu})_{23}|$ (Fig. 6).

 The term $\propto \Delta_{31}$ in eq. (\ref{23R})
can be neglected if, e.g.,  at least one of the 
CPV parameters $|a~\sin(\alpha/2)|$, 
$|b~\sin(\beta_M/2)\cos2\theta_{23}|$ 
($|b~\cos(\beta_M/2)|^2$) and 
$|c~\sin((\alpha-\beta_M)/2)\cos2\theta_{23}|$ 
($|c~\cos((\alpha-\beta_M)/2)|^2$)
is much bigger than $|\deltaatm|/(4m^2)$
($|\deltaatm|^2/(4m^2)^2$). 
In this case eqs. (\ref{12RQD}) - (\ref{QDQ})
are also valid. We get particularly simple expressions for
$|(Y_{\nu}^{\dagger}Y_{\nu})_{ij}^{\rm QD}|$, $i \neq j$,
if the terms $\propto a$ in eqs. (\ref{12R}) - (\ref{23R})  
dominate:
%%%%%%%%%%%%%%%%%%%%%%%%%%%%%%%%%%%%
\begin{align}
\label{12RQDa}
|(Y_{\nu}^{\dagger}Y_{\nu})^{\rm QD}_{12}|&\cong
2~|a|~\frac{M_R}{v_u^2}~\meff_{\rm QD}~c_{23} \;,\\ 
\label{13RQDa}
|(Y_{\nu}^{\dagger}Y_{\nu})^{\rm QD}_{13}|&\cong
|(Y_{\nu}^{\dagger}Y_{\nu})^{\rm QD}_{12}|~\tan\theta_{23}\;,\\
\label{23RQDa}
|(Y_{\nu}^{\dagger}Y_{\nu})^{\rm QD}_{23}|&\cong
2~|a|~\frac{M_R}{v_u^2}~
\sqrt{m^2 - \meff_{\rm QD}^2}~c_{23}s_{23}\;.
\end{align}
%%%%%%%%%%%%%%%%%%%%%%%%%%%%%%%%%%%%%%%%%%%%%
%
Equations (\ref{12RQDa}) - (\ref{13RQDa}) are valid provided  
$|a| \gg {\rm max}(|b|,|c|,|\deltaatm|/(4m^2))$,
while eq. (\ref{23RQDa}) holds if 
$|a~\sin(\alpha/2)|\gg {\rm max}(|b|,|c|,|\deltaatm|/(4m^2))$.
For $|a| < 1$ and, e.g., $m \cong 0.1$ eV, 
the latter condition requires 
$|\sin(\alpha/2)|\cong 1$.
In these cases both 
$\text{BR}(\mu \to e + \gamma) \sim |a|^2\meff_{\rm QD}^2$ 
and $\text{BR}(\tau \to e + \gamma)\sim |a|^2\meff_{\rm QD}^2$,
while $\text{BR}(\tau \to \mu + \gamma)\sim  
|a|^2~(m^2 - \meff_{\rm QD}^2) \cong 
|a|^2~m^2\sin^22\theta_{12}\sin^2(\alpha/2)$. For the ratio of the
first two we get 
%%%%%%%%%%%%%%%%%%%%%%%%%%%%%%
\begin{align}
\text{R}(21/31) \cong \cot^{2}\theta_{23}\;,
\label{R2131QD}
\end{align}
%%%%%%%%%%%%%%%%%%%%%%%%%%%%%%
%
which should be compared with eqs. (\ref{R2131NH})
and (\ref{R2131IHa}).
The ratio of $\text{BR}(\mu \to e + \gamma)$
and $\text{BR}(\tau \to \mu + \gamma)$
is independent of the leptogenesis CPV parameter $a$.
Given $\theta_{12}$ and $\theta_{23}$,
it is determined by the Majorana phase $\alpha$: 
%%%%%%%%%%%%%%%%%%%%%%%%%%%%%%
\begin{align}
\text{R}(21/32) \cong 
\frac{\meff_{\rm QD}^2}{(m^2 - \meff_{\rm QD}^2)~s_{23}^2}
\cong 
\frac{1 - \sin^22\theta_{12}\sin^2(\alpha/2)}
{s_{23}^2~\sin^22\theta_{12}\sin^2(\alpha/2)}
\;.
\label{R2132QDa}
\end{align}
%%%%%%%%%%%%%%%%%%%%%%%%%%%%%%
%
We get similar results if the terms $\propto b$ 
($\propto c$) dominate
in eqs. (\ref{12R}) - (\ref{23R}), which in the case of
eq. (\ref{23R}) would be possible only if
the Majorana phase $\beta_M$
(phase difference $\alpha - \beta_M$) 
deviates significantly from $\pi$.
Now $|(Y_{\nu}^{\dagger}Y_{\nu})^{\rm QD}_{23}|$ depends on 
$\beta_M$ ($\alpha - \beta_M$). If, e.g, the terms 
$\propto c$ dominate we get:
$\text{R}(21/32) \cong s_{23}^2\tan^2\theta_{12}~
(1 - \sin^22\theta_{23}\sin^2((\alpha - \beta_M)/2))^{-1}$.

  Our results for the double ratios $\text{R}(21/31)$ and  
$\text{R}(21/32)$ are illustrated in Fig. 5.
%%%%%%%%%%%%%%%%%%%%%%%%%%%%%%%%%%%%%%%%%%%
%%%%%%%%%%%%%%%%%%%%%%%%%%%%%%%%%%%%%%%%%%% 
\vspace{-0.2cm}
%%%%%%%%%%%%%%%%%%%%%%%%%%%%%%%%%%%%%%%%%%%%
%
\section{\large{Conclusions}}
%
%%%%%%%%%%%%%%%%%%%%%%%%%%%%%%%%%%%%%%%%%%%%

\vspace{-0.3cm}
\hskip 1.0cm Working in the framework of the class of 
SUSY theories with see-saw
mechanism and soft SUSY breaking with flavour-universal 
boundary conditions at a scale $M_X>M_R$,
we have analysed  the dependence of the rates of 
lepton flavour violating (LFV) decays 
$\mu\rightarrow e + \gamma$, 
$\tau \rightarrow e + \gamma$ 
$\tau \rightarrow \mu + \gamma$ ($l_i \to l_j + \gamma$) 
and of their ratios, on the Majorana and Dirac 
CP-violation (CPV) phases in  the PMNS matrix $\pmns$,
$\alpha$, $\beta_M$ and $\delta$, and 
on the leptogenesis CP-violating (CPV) parameters. 
The case of quasi-degenerate in mass 
heavy RH neutrinos was investigated,
$M_1 \cong M_2\cong M_3 \equiv M_R$,
assuming that splitting between the masses
of the heavy neutrinos is sufficiently small, 
so that it has practically no effect on the
predictions for the $l_i \to l_j + \gamma$
decay rates. Results for the 
normal hierarchical (NH), 
inverted hierarchical (IH) and quasi-degenerate (QD)
light neutrino mass spectra
have been derived. The analysis was
performed under the condition of negligible 
renormalization group (RG) effects for the 
light neutrino masses $m_j$ and the mixing angles 
and CPV phases in $\pmns$ 
\footnote{It is well-known that 
in the class of SUSY theories considered,
this condition is satisfied in the cases of 
NH and IH light neutrino mass spectra;
it is fulfilled for the QD spectrum 
provided the SUSY parameter $\tan\beta$ 
is relatively small, $\tan\beta < 10$.}.
In the wide region of validity of eqs. (\ref{eq_ijg}) 
and (\ref{eq_ms}) in the relevant SUSY 
parameter space, 
the ratios of rates of the decays 
$\mu\rightarrow e + \gamma$ and 
$\tau \rightarrow e + \gamma$,
and $\mu\rightarrow e + \gamma$ and
$\tau \rightarrow \mu + \gamma$, 
are independent of the SUSY parameters -
they are determined by the neutrino masses ($m_j$) and 
mixing angles, Majorana and Dirac CPV phases and by the
leptogenesis CPV parameter(s). 
 For the matrix of neutrino Yukawa couplings, $\mathbf{Y_{\nu}}$ -
a basic quantity in the analysis performed,
we have used the orthogonal parametrisation \cite{Iba01}.
The latter proved to be the most convenient for the
purposes of our study \cite{PPY03}. 
In this parametrisation $\mathbf{Y_{\nu}}$ is expressed 
in terms of the light and heavy Majorana neutrino masses, 
$\pmns$, and an orthogonal matrix $\mathbf{R}$~\cite{Iba01}.
Leptogenesis can take place only if 
$\mathbf{R} \neq {\bf 1}$ 
(see, e.g., \cite{LeptoG1,PPRi106}).
In the case of quasi-degenerate in mass 
heavy Majorana neutrinos 
considered in the present 
article, the rates of the LFV decays 
$l_i \to l_j + \gamma$ of interest 
do not depend on the matrix 
$\mathbf{R} \neq {\bf 1}$ if 
$\mathbf{R}$ is real.
For complex matrix $\mathbf{R}$,
only the three leptogenesis CPV
real dimensionless 
parameters of $\mathbf{R}$,
$a$, $b$ and $c$ (eqs. (\ref{RPPY}) and (\ref{Aabc})),
enter into the expressions for
the $l_i \to l_j + \gamma$ 
decay branching ratios of interest \cite{PPY03},
$\text{BR}(l_i \to l_j + \gamma)$.
In our analysis we have assumed that
$|a|,|b|,|c| < 1$, as is suggested by 
the leptogenesis constraint derived 
for QD light neutrinos 
and negligible favour effects
\cite{PPY03}. In various
estimates we have considered values of 
$|a|,|b|,|c| \leq 0.3$.  
The case of real matrix $\mathbf{R}$
corresponds effectively to 
% $a=0$, $b=0$, $c=0$.
$a,b,c=0$.

 We have found that for NH (IH) spectrum
and negligible lightest neutrino mass $m_1$ ($m_3$),
the branching ratios 
$\text{BR}(l_i \to l_j + \gamma)$
depend, in general, on one Majorana 
and the Dirac CPV phases, $\alpha - \beta_M$
($\alpha$) and $\delta$, one leptogenesis CPV 
parameter, $c$ ($a$), on the CHOOZ angle
$\theta_{13}$ and on the mixing angles
and mass squared differences associated with solar and 
atmospheric neutrino oscillations,
$\theta_{12}$, $\deltasol$, and
$\theta_{23}$, $\deltaatm$.
The double ratios 
$\text{R}(21/31) \propto \text{BR}(\mu \to e + \gamma)/
\text{BR}(\tau \to e + \gamma)$ and 
$\text{R}(21/32) \propto \text{BR}(\mu \to e + \gamma)/
\text{BR}(\tau \to \mu + \gamma)$ (see eq. (\ref{DoubleR}))
are determined by these parameters.
The same Majorana phase $\alpha - \beta_M$ ($\alpha$) 
enters also into the NH (IH) expression for the 
effective Majorana mass in neutrinoless double beta 
($\betabeta$-) decay, $\mefff$
(eqs. (\ref{meffNH2}) and (\ref{meffIH1})). 
For the QD spectrum, 
% the branching ratios
$\text{BR}(l_i \to l_j + \gamma)$ depend, in general,
on the absolute neutrino mass $m$,
the three leptogenesis CPV parameters, $a$, $b$, $c$
and on the two Majorana phases $\alpha$ and $\beta_M$. 
For the IH and QD spectra, 
the phase $\alpha$ enters into the expressions for 
$\text{BR}(\mu(\tau) \to e + \gamma)$,
% and $\text{BR}(\tau \to e + \gamma)$, 
in particular, through the effective 
Majorana mass $\mefff$ 
(see eqs. (\ref{IHP}) and (\ref{QDP})).
Our results for the double ratios show that 
we can have 
$\text{R}(21/31) \sim 1$ or $
\text{R}(21/31) \ll 1$,
or else $\text{R}(21/31) \gg 1$
in the cases of NH and IH spectra, 
while for the QD spectrum
typically $\text{R}(21/31) \sim 1$.
In contrast, for the NH and IH spectra one always 
gets $\text{R}(21/32) < 1$; in most of the 
relevant parameter space 
$\text{R}(21/32) \ll 1$ holds.
For the QD spectrum, however, $\text{R}(21/32) \gtap 1$ 
is also possible.

 More specifically, we find that for the NH (IH) spectrum, 
$\text{BR}(\mu (\tau) \to e + \gamma)$
exhibit  significant dependence  
on the leptogenesis CPV parameter $c$ ($a$)
and on the Majorana CPV phase $\alpha - \beta_M$
($\alpha$) for $|c| \gtap 0.02$
($|a|\gtap 0.02$) and for any 
$s_{13}\ltap 0.1$ ($s_{13}\ltap 0.2$). 
In certain cases the dependence 
of $\text{BR}(\mu (\tau) \to e + \gamma)$
on the phase $\alpha - \beta_M$ ($\alpha$)
and/or the parameter $c$ ($a$) is dramatic.
More generally, the dependence of 
$\text{BR}(\mu (\tau) \to e + \gamma)$ 
on the Majorana phase can 
be noticeable only if the corresponding 
leptogenesis parameter is sufficiently large:
for $|c| \ll {\rm max} ((\deltaatm/\deltasol)^{\frac{1}{4}}s_{13},
0.5(\deltasol/\deltaatm)^{\frac{1}{4}})$ 
in the NH case, and $|a| \ll 
{\rm max}(s_{13}/(2\cos2\theta_{12}),
\deltasol \tan2\theta_{12}/(8|\deltaatm|)$ in the IH one,
both $c~(a)$ and the Majorana phase have practically 
no effect on $\text{BR}(\mu (\tau) \to e + \gamma)$.
Similarly, the CHOOZ angle $\theta_{13}$ 
and the Dirac phase $\delta$ can be relevant 
in the evaluation of $\text{BR}(\mu (\tau) \to e + \gamma)$ 
in the cases of NH and IH spectra
only if $s_{13}$ is large enough, i.e., 
if respectively, $s_{13}\gtap \sqrt{\deltasol}
\sin2\theta_{12}/(2\sqrt{\deltaatm|}) \cong 0.07$, 
and $s_{13}\gtap \deltasol
\sin2\theta_{12}/(2|\deltaatm|) \cong 8\times 10^{-3}$.
In the case of NH (IH) spectrum,
$\text{BR}(\tau \to \mu + \gamma)$
is practically independent of $s_{13} \ltap 0.2$;
the dependence of $\text{BR}(\tau \to \mu + \gamma)$ on 
the leptogenesis parameter $c$ ($a$) and the Majorana 
phase $\alpha - \beta_M$ ($\alpha$)
is relatively weak for $|c| \ltap 0.3$ ($|a| \ltap 0.1$).
For this wide range of values of $|c|$ ($|a|$)
we have $\text{BR}(\tau \to \mu + \gamma) \cong 
\text{F} \times (|\deltaatm|/(4v^2_u))~
\sin^22\theta_{23}$, where the factor 
$\text{F} \propto M_R^2/v_u^2$ 
contains all the dependence on $M_R$, $\tan\beta$ and 
the SUSY breaking parameters (see eq. (\ref{eq_ijg})).

 The double ratios $\text{R}(21/31)$ and $\text{R}(21/32)$
(Figs. 1 - 4) can exhibit in the cases of NH and IH spectra
strong dependence on the Dirac and/or Majorana
phases if $s_{13} \sim 0.1 - 0.2$ 
and/or if the relevant leptogenesis 
parameter exceeds approximately 
$10^{-2}$. Under the indicated conditions values of  
$\text{R}(21/31) \sim (10^{-3} - 10^{-2}) \ll 1$ or
$\text{R}(21/31) \sim (10^{3} - 10^{2}) \gg 1$,
are possible. For, e.g., $s_{13} \sim 0.1$,  
the sign of the inequality is 
determined by the sign of 
the leptogenesis parameter,
the value of the Majorana phase
and/or the value of the Dirac phase (Figs. 1 and 3).
If for the  NH (IH) spectrum,
$\alpha - \beta_M\cong 0$ ($\alpha \cong \pi$) and
$\delta \cong \pm \pi/2,\pm 3\pi/2$,
$\text{R}(21/31)$ takes one of the following three values
$\text{R}(21/31) \cong 1;\tan^2\theta_{23};\cot^2\theta_{23}$.
For $|a| \gg \deltasol\tan2\theta_{12}/(8|\deltaatm|)$
in the IH case, we find $\text{BR}(\mu(\tau) \to e + \gamma)\cong
F^{\rm IH}\times|a|^2~|\mefff_{\rm IH}|^2/v_u^2$, and thus
$\text{R}(21/31) \cong 1$, where $\mefff_{\rm IH}$ is the  
effective Majorana mass in $\betabeta$-decay 
and the factor $F^{\rm IH} \propto M_R^2/v_u^2$ 
includes the dependence on $M_R$ and on the SUSY parameters. 
For sufficiently small $s_{13}$ and $|c|$  
($s_{13} \ll 0.07$, $|c| \ll 0.2)$) 
in the case of NH spectrum, we get: $\text{R}(21/32)\cong 
\deltasol/(\deltaatm s_{23}^2)~c^2_{12}s^2_{12} 
\cong 10^{-2}$. Smaller values of   
$\text{R}(21/32)$ are possible, e.g., for
$s_{13}\cong (0.1 - 0.2)$, if 
$|c| \sim 0.05$ and if for given ${\rm sgn}(c)$,
the Majorana and Dirac phases
$(\alpha - \beta_M)$ and $\delta$
have specific values (Fig. 2).
For the IH spectrum we 
typically have $\text{R}(21/32) \ll 1$ for
$|a|\leq 0.1$.
If $2|a|\cos2\theta_{12},\sin\theta_{13} 
\ll 0.5\deltasol~c_{12}s_{12}/|\deltaatm|$,
$\text{R}(21/32)$ is completely determined 
by the solar and atmospheric neutrino oscillation parameters 
$\deltasol$, $\theta_{12}$,  $\deltaatm$ and $\theta_{23}$, 
and $\text{R}(21/32) \cong 10^{-4}$.

 In the case of QD light neutrino mass spectrum, 
the leptogenesis constraint
implies \cite{PPY03} $10^{-6} \ltap |abc| \ltap 10^{-4}$.
The expressions for $\text{BR}(l_i \to l_j + \gamma)$
and for the double ratios 
$\text{R}(21/31)$ and $\text{R}(21/32)$ simplify
considerably if the terms including
one given leptogenesis parameter dominate.
We get, e.g., $\text{R}(21/31) \cong \tan^2\theta_{23}$
and $\text{R}(21/32) \cong 1$ if the terms 
$\propto b$ ($\propto c$) are the dominant one. 
This requires relatively large values of
$|a|$ or $|b|$  or $|c|$. If, however,
$|a| \sim |b| \sim |c| \simeq 10^{-2}$,
$\text{R}(21/32)$ lies in the interval
$\sim (10^{-4} - 10^{-1})$. 

\vspace{0.1cm}
{\bf Acknowledgments.} 
We would like to thank Y. Takanishi and W. Rodejohann
for useful discussions and T. Schwetz for 
informing us about results of analysis
including the MINOS data prior publication.
This work was supported in part by the Italian MIUR and INFN
programs on ``Fisica Astroparticellare''
and by the the European Network of Theoretical Astroparticle Physics 
ILIAS/N6 under the contract RII3-CT-2004-506222
(S.T.P.).

\vspace{0.1cm}
{\bf Note Added.} During the completion of the present study
we became aware \cite{WurzQD06} that an analysis along seemingly 
similar lines is being performed by R. R\"uckl et al.

%%%%%%%%%%%%%%%%%%%%%%%%%%%%%%%%%%%%%%%%%%%%%%%% 
% \newpage
%%%%% Added by T. SHINDOU%%%%%%%%%%%%%%%%%%%%%%%%
% \newpage
%%%%%%%%%%%%%%%%%%%%%%%%%%%%%%%%%%%%%%%%%%%%
\begin{figure}
\begin{tabular}{cc}
\includegraphics[scale=1.2]{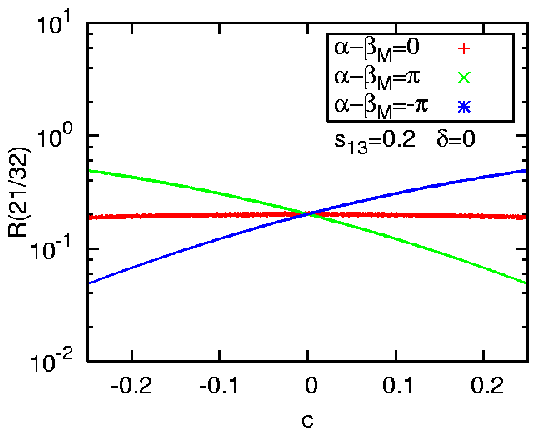}&
\includegraphics[scale=1.2]{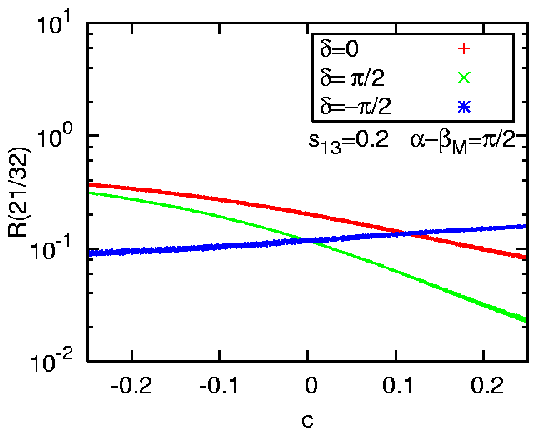}\\
(a)&(b)\\
\includegraphics[scale=1.2]{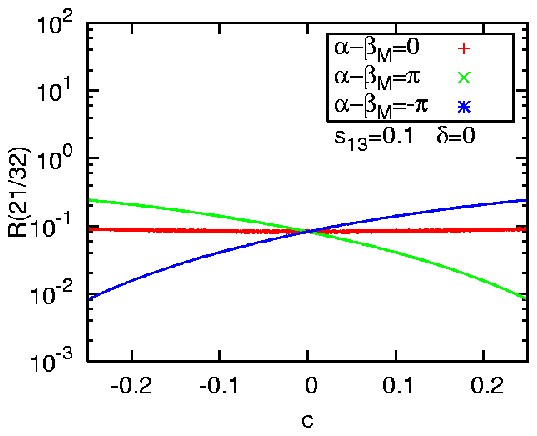}&
\includegraphics[scale=1.2]{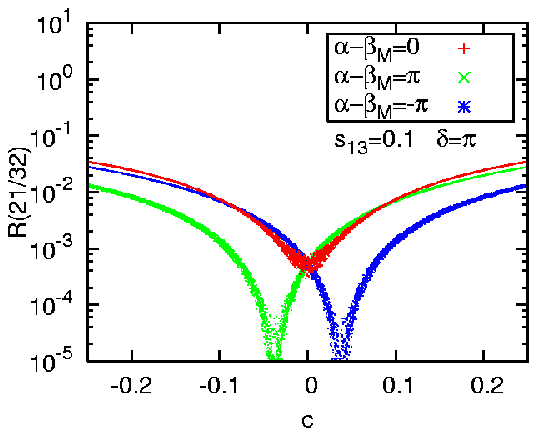}\\
(c)&(d)\\
\includegraphics[scale=1.2]{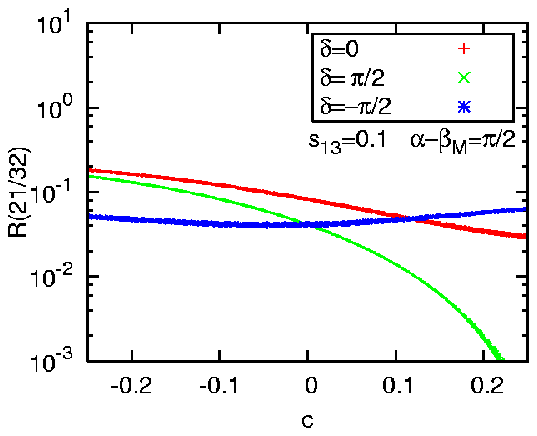}&
\includegraphics[scale=1.2]{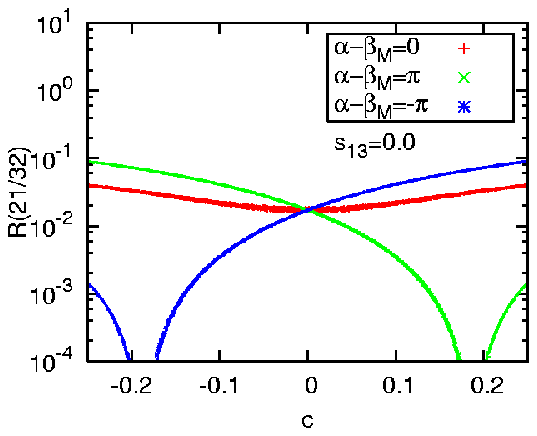}\\
(e)&(f)\\
\end{tabular}
\caption{The same as in Fig. \ref{NH-R2131-alp}, but 
for the double ratio $\text{R}(21/32)$.
}
\label{NH-R2132-alp}
\end{figure}
%%%%% end %%%%%%%%%%%%%%%%%%%%%%%%%%%%%%%%%%%%%%%%%%
\begin{figure}
\begin{tabular}{cc}
\includegraphics[scale=1.2]{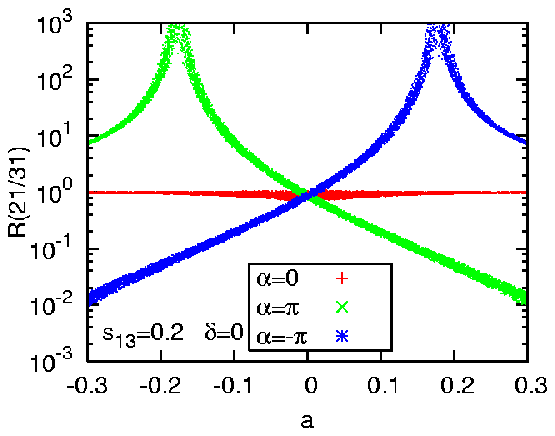}&
\includegraphics[scale=1.2]{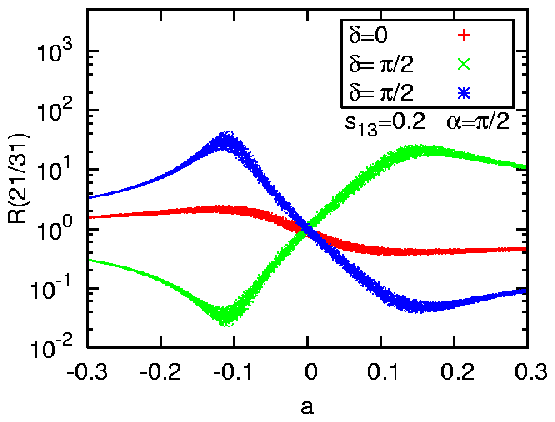}\\
(a)&(b)\\
\includegraphics[scale=1.2]{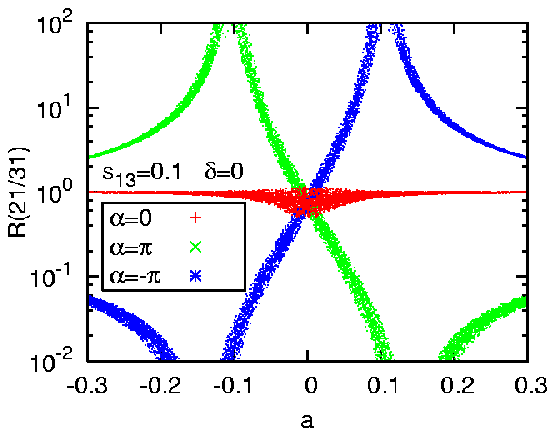}&
\includegraphics[scale=1.2]{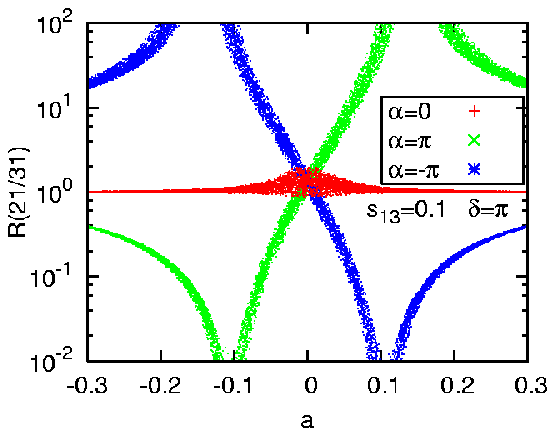}\\
(c)&(d)\\
\includegraphics[scale=1.2]{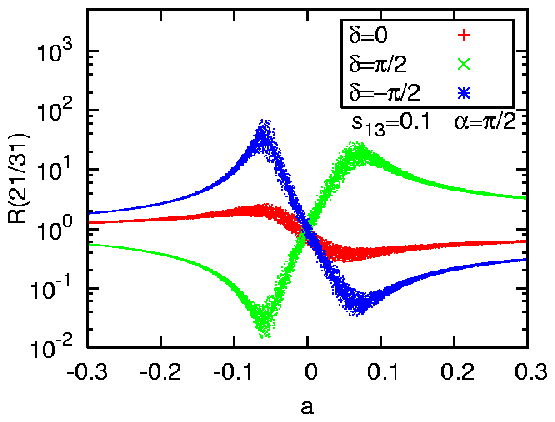}&
\includegraphics[scale=1.2]{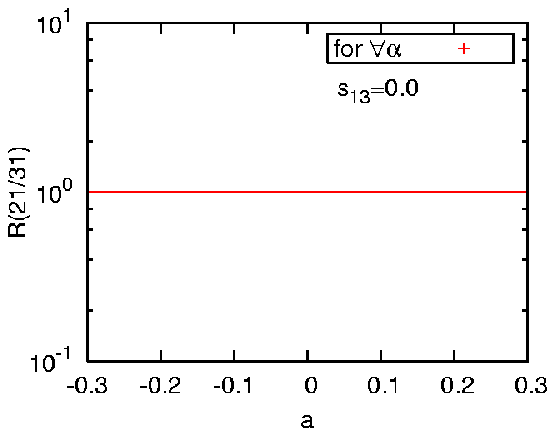}\\
(e)&(f)\\
\end{tabular}
\caption{The double ratio $\text{R}(21/31)$ % R(21/31) 
in the case of IH light neutrino mass spectrum
as a function of the leptogenesis CPV parameter $a$, 
for  several characteristic values of
the CHOOZ angle $\theta_{13}$ and Majorana and Dirac CPV phases 
$\alpha$ and $\delta$. The results shown correspond to 
the lightest neutrino mass $m_3 = 0$.
The effects of the higher order corrections 
due to the leptogenesis CPV parameters 
$b$ and $c$ is reflected in the widths of the lines
(see text for further details).
}
\label{IH-R2131-alp}
\end{figure}
%%%%%%%%%%%%%%%%%%%%%%%%%%%%%%%%%%%%%%%%%%%%%%%%%%
\begin{figure}
\begin{tabular}{cc}
\includegraphics[scale=1.2]{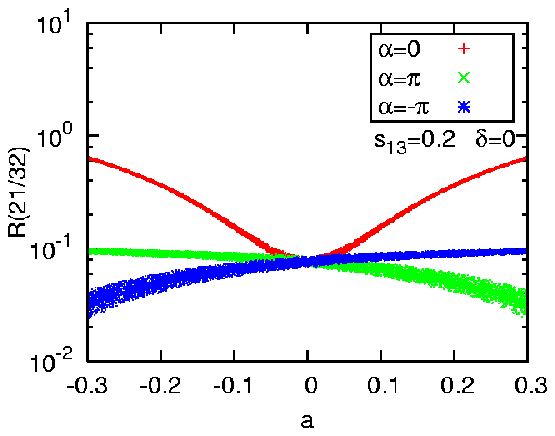}&
\includegraphics[scale=1.2]{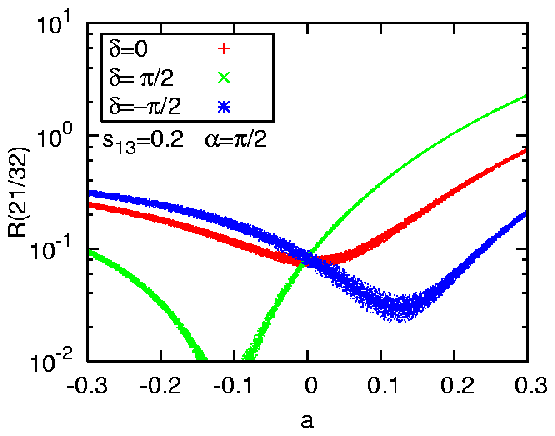}\\
(a)&(b)\\
\includegraphics[scale=1.2]{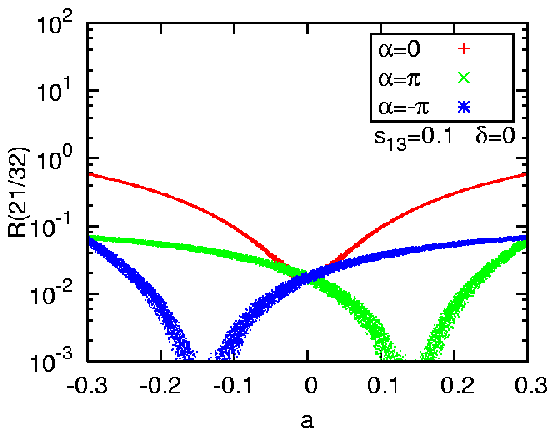}&
\includegraphics[scale=1.2]{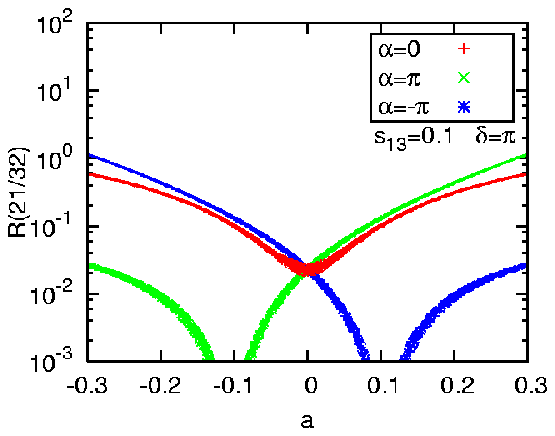}\\
(c)&(d)\\
\includegraphics[scale=1.2]{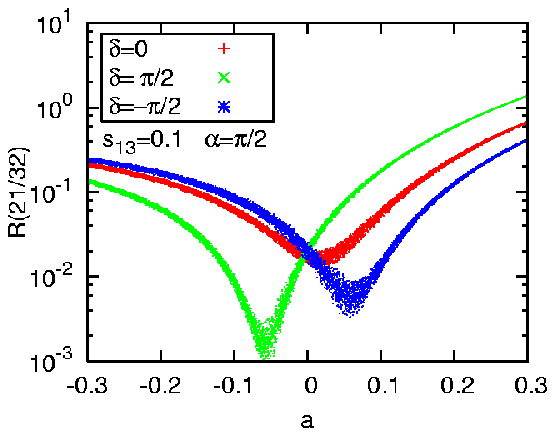}&
\includegraphics[scale=1.2]{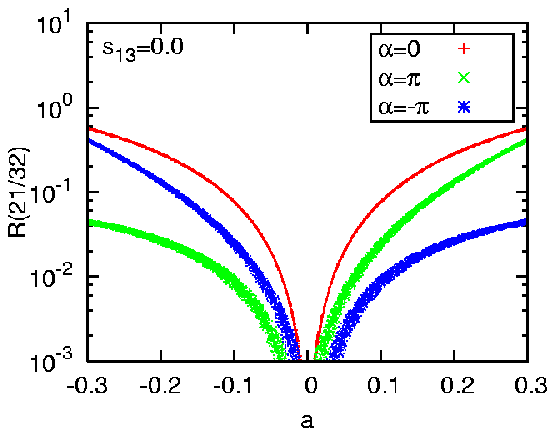}\\
(e)&(f)\\
\end{tabular}
\caption{The same as in Fig. 3, but for the double ratio
$\text{R}(21/32)$.
}
\label{IH-R2132-alp}
\end{figure}
\begin{figure}
\begin{tabular}{cc}
\includegraphics[scale=1.2]{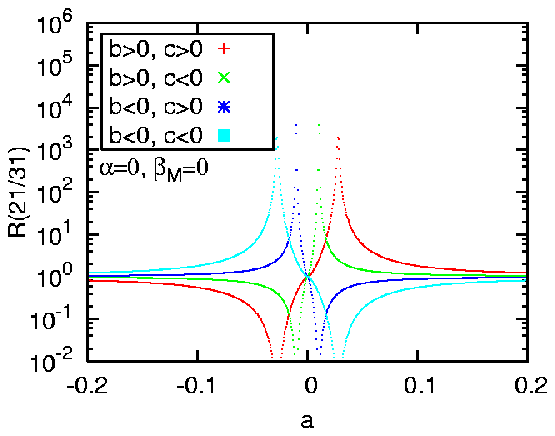}&
\includegraphics[scale=1.2]{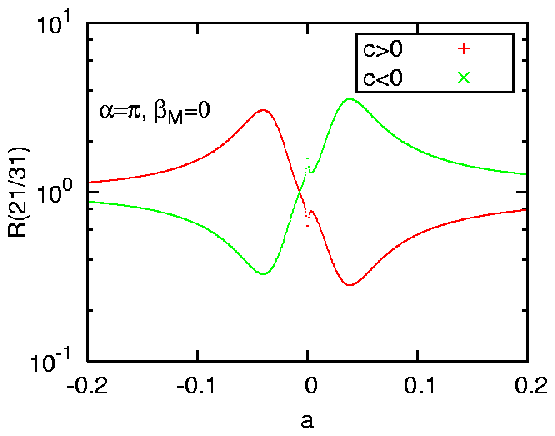}\\
\multicolumn{2}{c}{$s_{13}=0.0$}\\
\includegraphics[scale=1.2]{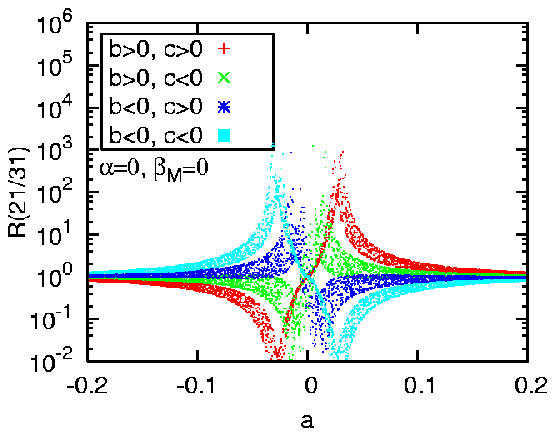}&
\includegraphics[scale=1.2]{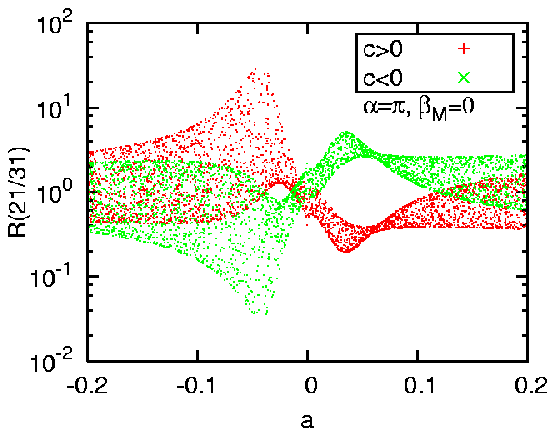}\\
\multicolumn{2}{c}{$s_{13}=0.1$}\\
\includegraphics[scale=1.2]{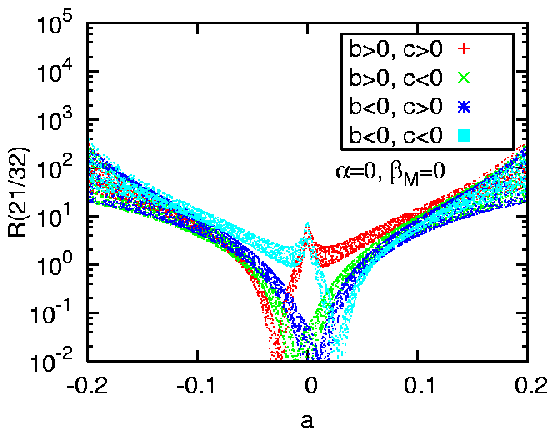}&
\includegraphics[scale=1.2]{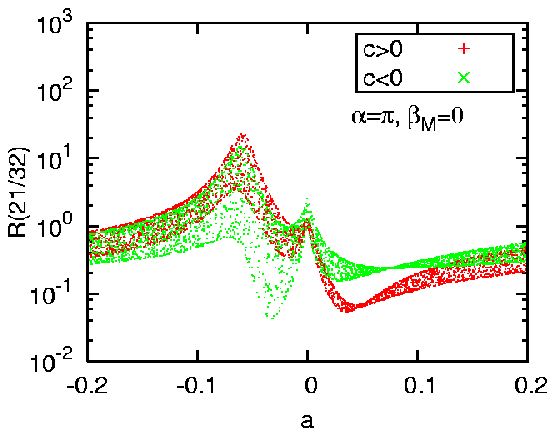}\\
\multicolumn{2}{c}{$s_{13}=0.1$}\\
\end{tabular}
\caption{The double ratios R(21/31) and R(21/32) as a function of the 
leptogenesis CPV parameter $a$ 
in the case of QD light neutrino mass spectrum 
for $s_{13}=0;0.1$ and several values of the Majorana phases
$\alpha$ and $\beta_M$.
The results shown are obtained for $|abc|=10^{-5}$ and $|b|=|c|$.
For $s_{13}=0.1$, values of the Dirac CPV phase
$0\leq \delta\leq \pi$ were considered.
The lightest neutrino mass is set to $m_1=0.1$ eV.
}
\label{QD-R2131}
\end{figure}
\end{document}